\documentclass[12pt,preprint]{aastex}

\newcommand{\der}[2][\;\;]{\ensuremath{ \frac{d{#1}}{d{#2}} }}
\newcommand{\dpar}[2][\;\;]{\ensuremath{ \frac{\partial{#1}}{\partial{#2}} }}
\newcommand{\dparn}[3][\;\;]{\ensuremath{ \frac{\partial^{#3}{#1}}{\partial{#2}^{#3}} }}

\newcommand{\e}{{\rm ext}}

\newcommand{\bvec}[1]{{\mbox{{\boldmath$#1$}}}} 

\newcommand{\eqnref}[1]{(\ref{#1})}

\begin{document}

\title{Axisymmetric Scattering of $p$ Modes by Thin Magnetic Tubes}

\author{Bradley W. Hindman}
\affil{JILA and Department of Astrophysical and Planetary Sciences,
University of Colorado, Boulder, CO~80309-0440, USA}

\author{Rekha Jain}
\affil{Applied Mathematics Department, University of Sheffield, Sheffield S3 7RH, UK}
\email{hindman@solarz.colorado.edu}


\begin{abstract}

We examine the scattering of acoustic $p$-mode waves from a thin magnetic fibril
embedded in a gravitationally stratified atmosphere. The scattering is mediated
through the excitation of slow sausage waves on the magnetic tube, and only the
scattering of the monopole component of the wavefield is considered. Since such
tube waves are not confined by the acoustic cavity and may freely propagate along
the field lines removing energy from the acoustic wavefield, the excitation of
fibril oscillations is a source of acoustic wave absorption as well as scattering.
We compute the mode mixing that is achieved and the absorption coefficients and
phase shifts. We find that for thin tubes the mode mixing is weak and the absorption
coefficient is small and is a smooth function of frequency over the physically
relevant band of observed frequencies. The prominent absorption resonances seen
in previous studies of unstratified tubes are absent. Despite the relatively small
absorption, the phase shift induced can be surprisingly large, reaching values
as high as 15$^\circ$ for $f$ modes. Further, the phase shift can be positive
or negative depending on the incident mode order and the frequency.

\end{abstract}



\section{Introduction}
\label{sec:introduction}
\setcounter{equation}{0}

It is well established that sunspots and magnetic plage are strong absorbers and
scatterers of the sun's acoustic $p$-mode waves \citep[e.g.,][]{Braun:1995, Braun:2008,
Haber:1999}. The absorption is unexpectedly large, reaching values as high as 70\%
for sunspot groups and a lesser, but still substantial, value of 20\% for magnetic
plage. Equally strong phase shifts are also observed for sunspots, with shifts as
large at 100$^\circ$ being possible. Interestingly, plages lack measurable phase
shifts, despite possessing significant absorption \citep{Braun:1995}.

Initial attempts to model the interaction of acoustic waves with magnetic features
concentrated on the absorption caused by sunspots, as the observed signal was strong
and easy to measure. A host of absorption mechanisms were investigated, including
resonant absorption \citep[e.g.,][]{Hollweg:1988, Lou:1990, Keppens:1994}, mode
mixing \citep{D'Silva:1994} and mode conversion \citep[e.g.,][]{Cally:1993, Cally:2000}.
The present belief is that mode conversion from acoustic waves to slow magnetosonic
waves within an appreciably inclined magnetic field is responsible \citep{Crouch:2003}.
Under this paradigm, acoustic energy is transformed into slow waves which propagate
along the field lines removing energy from the $p$-mode cavity. The conversion
process is most efficient when the phase velocity of the incident acoustic wave
matches the local phase velocity of the slow mode. Such strong coupling of the
phase velocities occurs when two conditions are met: the wavenumber vector of the incident
acoustic wave is nearly parallel to the magnetic field and the Alfv\'en and sound
speeds are comparable. In a sunspot, the Alfv\'en and sound speeds are equal to
each other in a layer just below the solar surface and the ray paths of the $p$ modes
in this layer are such that the strongest conversion occurs in regions with field
lines that are inclined 20--30 degrees from vertical (such as within penumbrae).
In addition to successfully explaining the absorption observed for sunspots, mode
conversion models have also succeeded in reproducing the large phase shifts
\citep{Cally:2003}.

So, while we presently have a fairly good understanding of how a flaring, monolithic,
magnetic structure such as a sunspot can absorb and scatter acoustic waves, we have made
little progress in modelling the observed absorption by plage. The absorption within
a plage is smaller than that of a sunspot by only a factor of two, yet the mode
conversion process is unlikely to work in the same manner, as the field within a plage
has a fractured and fibril nature, and probably has less inclined field of a structured
geometry in the photospheric layers. The limited theoretical work that has been done
modelling the effects of fibril magnetic fields indicates that multiple scattering may
play a very important role \citep{Bogdan:1987, Keppens:1994, Bogdan:1991}. \cite{Keppens:1994}
examined multiple scattering between a small number of flux tubes (2--3) in an unstratified
atmosphere and found that resonant absorption by the collection of tubes can be greatly
enhanced if the tubes are tightly spaced. \cite{Tirry:2000} performed a similar calculation
but considered incident $p$-mode waves trapped within a solar-like wave guide and their
interaction with thin flux tubes. Tirry permitted tube waves excited on the magnetic
fibrils to freely cross the bottom boundary of the acoustic waveguide and discovered
that such wave leakage destroyed the absorption resonances observed by \cite{Keppens:1994}.
Both studies lacked gravitational stratification. Neither of these studies included a
sufficiently large number of tubes to reliably estimate the expected phase shifts measured
in plage. The lack of measurable phase shift is likely to be the result of incoherent
scattering from a large number of tubes.

\cite{Jain:2009} examined whether the wave leakage mechanism explored by Tirry is
sufficiently large on its own to explain the absorption observed within plage. They
modeled a plage as a large collection of thin magnetic tubes embedded in a stratified
atmosphere and considered each tube to absorb acoustic waves in isolation from the other
members of the collective. The MHD tube waves excited on each tube by the incident $p$
mode freely propagate along the fibril out of the acoustic cavity despite the gravitational
stratification. They found that high levels of absorption (potentially greater than 50\%)
were easily achieved. However, they did not take into account scattering between tubes
or the screening that is likely to occur for tubes located far from the edge of the plage. 

To date few studies have examined both the {\sl absorption} and {\sl scattering}
of acoustic waves by magnetic tubes residing in a {\sl gravitationally stratified}
atmsophere. Notable are those of \cite{Hanasoge:2008} and \cite{Hanasoge:2009} which
focused on the scattering of the dipole component of incident $f$ modes from thin
magnetic tubes through the excitation of kink waves, and the study by \cite{Gordovskyy:2009}
that, through numerical simulation, examined axisymmetric scattering from tubes
of general thickness. Here we examine scattering and absorption of the wavefield's
monopole component of both $f$ and $p$ modes from a thin magnetic tube in a gravitationally
stratified atmosphere. The scattering is mediated through the excitation of axisymmetric
sausage waves that travel up and down the fibril.

The paper is organized as follows: in Section \ref{sec:formalism} we present our
scattering formalism. In Section \ref{sec:matrices}, we calculate the scattering
matrices. Section \ref{sec:absorption} describes the absorption achieved and the
properties of the mode mixing across mode order that occurs, while Section \ref{sec:phases}
presents the resulting phase shifts. Finally, in Section \ref{sec:discussion} we
discuss our findings and conclusions, including what our results for a single tube
adumbrate for plage.


\section{Scattering Formalism}
\label{sec:formalism}
\setcounter{equation}{0}

We consider a compact magnetic fibril that is embedded within a field-free atmosphere.
The fibril is axisymmetric, straight and aligned with gravity. Under such conditions
the acoustic wavefield in the surrounding, nonmagnetized atmosphere can be expressed
as the sum of three components,

\begin{equation} \label{eqn:total_Phi}
    \Phi(\bvec{x},t) = \Phi_{\rm inc}(\bvec{x},t) + \Phi_{\rm sca}(\bvec{x},t) +
                                \Phi_{\rm jac}(\bvec{x},t) \; .
\end{equation}

\noindent Here $\Phi$ can represent any physical variable we might choose to consider;
however, we will restrict our attention to a plane-parallel atmospheric model that is
neutrally stable to convection---a good approximation within the sun's convection zone.
For such an atmosphere, acoustic waves can be described using a displacement potential
and we will let $\Phi$ be this potential. The first term, $\Phi_{\rm inc}$, is the
unperturbed $p$-mode wavefield that would exist in the absence of the magnetic flux
concentration (i.e., the {\sl incident} wavefield). The second term, $\Phi_{\rm sca}$
represents a discrete set of far-field {\sl scattered} waves in the form of propagating
$p$ modes.  Finally, $\Phi_{\rm jac}$ is the contribution due to the continuous spectrum
of laterally evanescent, near-field, {\sl jacket} modes \citep[see][]{Bogdan:1995}.

We can express an arbitrary wavefield of incident $p$ modes as a Fourier-Bessel
expansion in cylindrical polar coordinates, $\bvec{x} = (r,\phi,z)$, where the origin
is centered on the flux concentration,

\begin{equation} \label{eqn:incident}
    \Phi_{\rm inc}(\omega;r,\phi,z) = \sum_{m = -\infty}^\infty \sum_{n=0}^\infty
                {\cal A}_{mn}(\omega) \; e^{im\phi} \; J_m(k_n r) \; Q_n(\omega;z) \; .
\end{equation}

\noindent In this expression---and in the subsequent two---we have chosen to apply
temporal Fourier transforms $(t\to\omega)$ and to work in frequency space. Our Fourier
convention is such that our solutions have a temporal dependence of $e^{-i\omega t}$.
The coefficients ${\cal A}_{mn}(\omega)$ are arbitrary complex amplitudes that
characterize the incident wavefield, $J_m(k_n r)$ is the Bessel function of the first
kind, $Q_n(\omega;z)$ is the vertical eigenfunction for the $n$th-order $p$ mode, and
$k_n = k_n(\omega)$ is the wavenumber eigenvalue for the $n$th-order $p$ mode with
frequency $\omega$.

Since we have assumed that the magnetic fibril is axisymmetric, incident waves of
azimuthal order $m$ only scatter into waves with the same order. Therefore, for such
an axisymmetric scatterer, the scattering only occurs across mode order $n$, and the
explicit form of the scattered waves can be represented as follows:

\begin{eqnarray}
    \Phi_{\rm sca}(\omega;r,\phi,z) &=& \sum_{m = -\infty}^\infty \sum_{n=0}^\infty
                {\cal A}_{mn}(\omega) \; e^{im\phi} \sum_{n'=0}^\infty \;
                S_m^{n \to n'}(\omega) \; H_m^{(1)}(k_{n'}r) \; Q_{n'}(\omega;z) \; ,
    \label{eqn:scattered}                               \\
\nonumber                                       	\\
    \Phi_{\rm jac}(\omega;r,\phi,z) &=& \sum_{m = -\infty}^\infty \sum_{n=0}^\infty
                {\cal A}_{mn}(\omega) \; e^{im\phi} \int_{0}^\infty d\Lambda \;
                T_m^{n \to \Lambda}(\omega) \; K_m(\Lambda r) \; q(\omega, \Lambda; z) \; .
    \label{eqn:jacket}
\end{eqnarray}

\noindent In these expressions $H_m^{(1)}(k_{n'} r)$ is the Hankel function of
the first kind---corresponding to an outward propagating cylindrical wave, and
$K_m(\Lambda r)$ is the modified Bessel function of the second kind, with $\Lambda^{-1}$
being the lateral decay length for the jacket modes. There are two separate scattering
matrices: $S_m^{n\to n^\prime}$ is the far-field scattering matrix that represents
scattering into outgoing, propagating $p$ modes, whereas $T_m^{n \to \Lambda}$ is
the near-field matrix and represents the excitation of the acoustic jacket. In
this formalism, mode mixing appears as nonzero values for the off-diagonal elements
of the far-field scattering matrix (i.e., $S_m^{n \to n^\prime} \neq 0$ for
$n \neq n^\prime$). The functions $q(\omega,\Lambda;z)$ describe the vertical behavior
of each jacket mode and are oscillatory with depth. We note that the $p$ mode
eigenfunctions $Q_n(z)$ and the jacket modes $q(\Lambda;z)$ form a complete, orthogonal
set. We shall make use of this orthogonality property in the following section to
compute the scattering matrices. Appendix~\ref{app:eigenfunctions} describes in detail
the $p$ mode eigenfunctions and jacket waves appropriate for the neutrally-stable,
polytropic atmosphere.

In general, the two scattering matrices, $S_m^{n\to n'}$ and $T_m^{n\to \Lambda}$,
tell us everything we need to know about the wave-tube interaction, including the
absorption coefficient $\alpha_{mn}$ and phase shift $\Delta_{mn}$. For a particular
mode (i.e., a particular $m$ and $n$) the absorption coefficient is defined as one
minus the ratio of the outgoing power to the ingoing power contained by that mode,

\begin{equation} \label{eqn:abscoef_def}
	\alpha_{mn}(\omega) \equiv \frac{\left|A_{mn}^{\rm (in)}(\omega)\right|^2 -
		\left|A_{mn}^{\rm (out)}(\omega)\right|^2}{\left|A_{mn}^{\rm (in)}(\omega)\right|^2}
		= 1 - \left| \frac{A_{mn}^{\rm (out)}(\omega)}{A_{mn}^{\rm (in)}(\omega)}\right|^2.
\end{equation}

\noindent The phase shift is defined in the usual way, as the phase difference between
the ingoing and outgoing components,

\begin{equation} \label{eqn:phaseshift_def}
	\Delta_{mn}(\omega) \equiv \arg \left\{ \frac{A_{mn}^{\rm (out)}(\omega)}
					  { A_{mn}^{\rm (in)}(\omega)} \right\}.
\end{equation}

If we consider only a single incident wave, we can relate the absorption coefficient
and phase shift to the scattering matrices in a straight forward manner. Noting that
an incident wave component is comprised of outgoing and ingoing parts of equal amplitude,
$J_m = \left(H_m^{(1)} + H_m^{(2)}\right)/2$, the outgoing and ingoing amplitudes,
$A^{\rm  (out)}$ and $A^{\rm (in)}$, are then given by,

\begin{eqnarray}
	A_{mn}^{\rm (in)}(\omega) &=& \frac{{\cal A}_{mn}(\omega)}{2} ,
\\ \label{eqn:A_out}
	A_{mn}^{\rm (out)}(\omega) &=& \frac{{\cal A}_{mn}(\omega)}{2}
			\left[ 1 + 2 S_m^{n \to n}(\omega) \right] .
\end{eqnarray}

\noindent If we now insert these expressions for the ingoing and outgoing power into
equations~\eqnref{eqn:abscoef_def} and \eqnref{eqn:phaseshift_def}, we find

\begin{eqnarray}
    \alpha_{mn}(\omega) &=& 1 - \Big|1 + 2 S_m^{n \to n}(\omega)\Big|^2 \; ,	\\
    &=& -4 {\rm Re}\left\{S_m^{n \to n}\right\} - 4 \Big|S_m^{n \to n}\Big|^2 \; ,
    	\label{eqn:abscoef_singlemode}                         	\\
\nonumber							\\
    \Delta_{mn}(\omega) &=& \arg\left\{1 + 2 S_m^{n \to n}(\omega)\right\} \; ,	\\
    &=& \tan^{-1}\left(\frac{2{\rm Im}\left\{S_m^{n \to n}\right\}}
		{1+2{\rm Re}\left\{S_m^{n \to n}\right\}}\right) \; .
        \label{eqn:phaseshift_singlemode}
\end{eqnarray}

For realistic wave fields (e.g., as sampled by helioseismic observations) these simple formulae
are invalid. Firstly, multiple incident waves of all mode orders $n$ exist simultaneously.
Secondly, mode mixing (the scattering from one mode order $n$ to a different mode order
$n^\prime$) will result in the outgoing component of any mode order being comprised of the
scattered waves from all incident mode orders.
One goal of this paper is to assess the importance of mode mixing in the energy budget of
an incident wave. Fortunately this can be accomplished by examining each incident mode in
isolation, and computing the off-diagonal elements of the scattering matrix. We will leave
for a subsequent study the effects that a complex, multi-component incident wave field has
on measured absorption coefficients and phase shifts. Further, we will only consider mode
mixing across mode orders. We presume that the scatterer is axisymmetric; therefore, scattering
from one azimuthal order $m$ to another $m'$ does not occur. Finally, for simplicity we
have chosen to examine only the axisymmetric component ($m=0$) of the incident wavefield,
leaving higher-order waves to subsequent studies. The index $m$ will be dropped from most
expressions from here forward.


\section{Calculating the Scattering Matrices}
\label{sec:matrices}
\setcounter{equation}{0}

Scattering matrices have been computed previously for axisymmetric magnetized tubes
within unstratified atmospheres \citep[e.g.,][]{Keppens:1994, Tirry:2000}. However,
calculating the scattering matrices for a magnetic tube in a gravitationally stratified
atmosphere is formidable and can only be accomplished in general through numerical
wave-field simulations. However, if we assume that the magnetic flux tube is thin
and untwisted, substantial progress can be made without resorting to computer simulation.
By {\sl thin}, we mean that all characteristic scale lengths---such as the wavelength
of the external acoustic oscillations and the density scale height---are much larger
than the radius of the tube $R(z)$. Such a thin flux tube, although in hydrostatic
balance, is unable to support internal lateral forces; hence, the total pressure
will be uniform across the tube and continuous with the external value. This mandates
that the total pressure has the same scale height inside and outside the tube. If
we further assume that the temperature is the same within the tube as outside, the
plasma parameter $\beta$, defined as the ratio of the gas pressure to the magnetic
pressure, is constant with height inside the tube. For photospheric and deeper layers,
this assumption that the tube is in thermal equilibrium with its surroundings is
consistent with observations \citep{Lagg:2010}. Thin tubes with constant $\beta$
are theoretically attractive, as they permit analytic treatment of the waves that
propagate along them and prove to be weak scatterers for which asymptotic analysis
may be used.

The scattering matrices are determined by the application of two physical matching
conditions. At the interface between the magnetic tube and the surrounding field-free
atmosphere, both the total pressure perturbation and the normal component of the
fluid displacement must be continuous. In general, thin flux tubes permit two types
of magnetosonic waves, sausage and kink waves.  The sausage waves are axisymmetric ($m=0$),
longitudinal, slow surface waves with pressure as the primary restoring force. The kink
waves are sinuous, transverse, fast waves with $m = \pm 1$ and a restoring force composed
of magnetic tension and buoyancy. Since we will only be considering axisymmetric incident
waves, the solution within the magnetic tube will be composed of only sausage waves.
We mention for completeness that magnetic tubes also permit torsional Alfv\'en waves
which are also axisymmetric oscillations. However, acoustic-gravity waves within a
neutrally-stable atmosphere are irrotational and are unable to couple to torsional
waves.

The propagation of sausage waves on magnetic fibrils is reasonably well understood
\citep[e.g.,][]{Defouw:1976, Roberts:1978}. Such waves have the property that the total
pressure is nearly constant across the cross section of the tube and matches the overpressure
in the field-free surrounding media. Furthermore, the normal component of the fluid displacement
at the boundary of the tube is a small quantity of the same size as the radius of the
tube (which is presumed small by the thin tube approximation). The smallness of the normal
displacement enables the solution inside the tube to be obtained to leading order with
only the enforcement of pressure continuity; the continuity of the displacement appears
only in higher-order corrections \citep{Spruit:1979, Spruit:1982, Andries:2011}. Here we
explicitly consider these higher order terms in order to compute the scattered wave field. 

Outside the flux tube, a small argument expansion of the Bessel functions appearing in
equation~\eqnref{eqn:incident} reveals that, to lowest order in the radius of the flux
tube $R$, the pressure perturbation of the incident wave is a zero-order quantity ${\rm O}(R^0)$,
whereas the normal displacement is first order ${\rm O}(R)$. Thus, the ordering of terms
for the incident waves is the same as the sausage waves. However, similar expansions of
equations~\eqnref{eqn:scattered} and \eqnref{eqn:jacket} reveal that the far-field and
near-field scattered waves have a pressure perturbation whose leading-order terms are
proportional to $|S| \ln R$ or $|T| \ln R$, respectively, where $|S|$ and $|T|$ indicate
the leading-order behavior of the scattering matrices. The normal displacements for these
waves are proportional to $|S| R^{-1}$ and $|T| R^{-1}$. The size of each of these terms
for each wave component is summarized in Table 1. Careful scrutiny reveals that the manner
in which a balance can be successfully achieved, where both continuity conditions are satisfied,
is if the scattering matrices are to leading order proportional to $R^2$. With this behavior,
pressure continuity is maintained by a balance between the sausage wave and the incident wave;
the scattered waves play no role to leading order. Continuity of the normal displacement
involves a balance between the sausage wave and all three external wave components (incident,
far-field scattered and near-field scattered).

\bigskip
\centerline{\small Table 1: LEADING ORDER OF EACH WAVE COMPONENT}

\begin{center}
{\footnotesize
\begin{tabular}{ccc}

{Wave Component} & {Pressure Perturbation} & {Normal Displacement}\\
\hline
\hline
External Incident Wave  & $R^0$          & $R$          \\
External Scattered Wave & $|S|\ln R$ & $|S|R^{-1}$ 	\\
External Jacket Wave    & $|T|\ln R$ & $|T|R^{-1}$ 	\\
\hline
Internal Sausage Wave   & $R^0$          & $R$          \\
\hline
\hline

\end{tabular}

}
\end{center}

By these arguments, the sausage waves are excited by the incident wave through
pressure continuity and the scattered waves (and the scattering matrices) are
generated by the sausage waves through continuity of the normal component of
the displacement. The normal displacement outside the tube $N_\e$ is a linear
combination of wave components,

\begin{equation} \label{eqn:Next}
	N_\e(\omega;z) = \sum_{n=0}^\infty \frac{{\cal A}_{n}(\omega)}{z_0^2}
		\left[N_{{\rm inc},n}(\omega;z) + N_{{\rm sca},n}(\omega;z) + N_{{\rm jac},n}(\omega;z)\right] \; ,
\end{equation}

\noindent where $N_{{\rm inc},n}$ is the normal displacement for the incident wave
of order $n$ and $z_0$ is a scale depth (chosen to be the depth of the photosphere
in our field-free atmospheric model---see Appendix~\ref{app:eigenfunctions}). The
displacements $N_{{\rm sca},n}$ and $N_{{\rm jac},n}$
are the far-field and near-field scattered waves generated by incident mode order $n$.
All displacements are evaluated at the flux-tube interface $r = R(z)$ and any given
displacement can be computed from the associated displacement potential through the
relation,

\begin{equation} \label{eqn:Ncomp}
	N = \bvec{\hat{n}} \cdot \bvec{\nabla}\Phi = \left(\dpar[\Phi]{r}-\der[R]{z}\dpar[\Phi]{z}\right)
			\left[1+\left(\der[R]{z}\right)^2\right]^{-1} \; ,
\end{equation}

\noindent where $\bvec{\hat{n}}$ is the unit vector normal to the surface. The
leading order behavior of each of these terms is found by small argument expansion
of the Bessel functions in equations~\eqnref{eqn:incident}--\eqnref{eqn:jacket},

\begin{eqnarray}
    N_{{\rm inc},n}(\omega;z) &=& -R(z) z_0^2 \left[\frac{k_n^2}{2} \; Q_n(\omega;z)
        + \frac{1}{R}\frac{dR}{dz} \; \frac{dQ_n}{dz}(\omega;z) \right] \; ,
    \label{eqn:Ninc}                                \\
\nonumber                                           \\
    N_{{\rm sca},n}(\omega;z) &=& \frac{z_0^2}{R(z)} \; \frac{2i}{\pi} \sum_{n'=0}^\infty \;
                S_0^{n\to n'}(\omega) \; Q_{n'}(\omega;z) \; ,
    \label{eqn:Nsca}                                 \\
\nonumber                                           \\
    N_{{\rm jac},n}(\omega;z) &=& - \frac{z_0^2}{R(z)} \; \int_{0}^\infty d\Lambda \;
                T_0^{n\to \Lambda}(\omega) \; q(\omega,\Lambda;z) \; .
    \label{eqn:Njac}
\end{eqnarray}

Each order of incident wave excites a separate sausage wave on the magnetic fibril,

\begin{equation} \label{eqn:Ntube}
	N_\parallel(\omega;z) = \sum_{n=0}^\infty \frac{{\cal A}_{n}(\omega)}{z_0^2} N_{\parallel,n}(\omega;z) \; .
\end{equation}

\noindent Appendix~\ref{app:sausagewaves} provides details on how to compute each
$N_{\parallel,n}(\omega;z)$. Continuity of the normal displacement requires that
$N_{\parallel,n} = N_{{\rm inc},n} + N_{{\rm sca},n} + N_{{\rm jac},n}$. If we insert
equations~\eqnref{eqn:Ninc}---\eqnref{eqn:Njac} into this relationship and multiply
by $R/z_0^2$, we obtain an equation for the ``mismatch" between the incident and tube
waves,

\begin{eqnarray}
        d_n(\omega; z) &\equiv& \frac{R(z)}{z_0^2}
		\left[N_{{\rm inc},n}(\omega;z) - N_{\parallel,n}(\omega;z) \right] \; , 
	\label{eqn:dn}				\\
	&=& -\frac{2i}{\pi} \sum_{n'=0}^\infty S_0^{n \to n'}(\omega) \; Q_{n'}(\omega;z)
                + \int_0^\infty d\Lambda \; T_0^{n \to \Lambda}(\omega) \; q(\omega, \Lambda; z) \; .	
\end{eqnarray}

\noindent Figure~\ref{fig:wave_functions}  illustrates the two displacement components for
an incident $p_2$ mode, the components for the resulting sausage wave, and the mismatch
between them $d_n$.

We can now solve for the two scattering matrices separately by invoking the orthogonality
of the eigenfunctions and the jacket waves. In Appendix~\ref{app:eigenfunctions} we
discuss that the weighting function for the orthogonality integrals is proportional
to the mass density. If we make use of a dimensionless depth $s = -z/z_0$, the density
within a neutrally stable atmosphere is a power law $\rho = \rho_0 s^a$, where $\rho_0$
is the photospheric density and the polytropic index $a$ is related to the ratio of specific
heats $\gamma$ through $a=(\gamma-1)^{-1}$. Multiply equation~\eqnref{eqn:dn} by $s^a$
and the complex conjugate of either a $p$-mode eigenfunction $Q$ or a jacket wave $q$ and
integrate over all depths to obtain,

\begin{eqnarray}
    S_0^{n \to n'}(\omega) &=& \frac{i\pi}{2}\int_1^\infty ds \; s^a \; Q_{n'}(\omega;s) \; d_n(\omega;s) \; ,
    \label{eqn:Smatrix}                                     \\
    T_0^{n \to \Lambda}(\omega) &=& \int_1^\infty ds \; s^a \; q^*(\omega,\Lambda;s) \ d_n(\omega;s) \; .
    \label{eqn:Tmatrix}
\end{eqnarray}

\noindent The limits of integration correspond to the photosphere ($s=1$) and infinitely
deep into the solar interior.

Each element of the scattering matrices can therefore be thought of as the projection
of the mismatch between the incident and scattered waves onto each $p$-mode and jacket-mode
eigenfunction. The matrix associated with the far-field scattering, $S_0^{n \to n'}(\omega)$
can be easily obtained by direct numerical integration of equation~\eqnref{eqn:Smatrix}.
We have used a Bulirsch-Stoer numerical integrator to perform these integrations and the
results are discussed in the following sections. The integral that must be computed to
obtain $T_0^{n \to \Lambda}(\omega)$ is fundamentally more difficult to numerically
compute since jacket wavefunctions are not confined to an acoustic cavity and remain
propagating to any depth. Therefore, while the magnitude of the integrand does vanish
as $s \to \infty$, it does so slowly as a power law. This is unlike the integrand of
Equation~\eqnref{eqn:Smatrix} which behaves like the product of a power law times a
rapidly decaying exponential as $s \to \infty$. Therefore, we delay the calculation
of the near-field scattering matrix to a subsequent time. Fortunately, the far-field
and near-field scattering matrices are independent of each other due to the orthogonality
of the $p$ modes and the jacket modes; thus, using this technique to calculate the
far-field scattering matrix avoids the necessity of computing the acoustic jacket
simultaneously.

Under many reasonable boundary conditions, one can demonstrate that the far-field
scattering matrix $S_0^{n \to n'}$ is symmetric under exchange of $n$ and $n'$ (see
Appendix~\ref{app:SymmetryS} for a derivation). Here we apply a stress-free condition
at the photosphere (for both the $p$ modes and the sausage waves) and finite energy
(for the $p$ modes) and radiation (for the sausage waves) boundary conditions deep
in the atmosphere (as $s \to \infty$). These boundary conditions are among those for
which the scattering matrix is symmetric. This means that we need not calculate all
of the matrix elements, only a little more than half need to be computed directly
with the remaining obtained by symmetry.


\section{Absorption and Mode Mixing}
\label{sec:absorption}
\setcounter{equation}{0}

The absorption coefficient is often considered to be a measurement of energy loss
or dissipation. When mode mixing occurs, this simple scenario needs to be
modified as in addition to true absorption (or removal of energy from the acoustic
wavefield), energy can be shuffled between modes by mode mixing. In this section
we calculate the energy budget for a single incident wave and account for all the
ways that the incident wave energy can be reprocessed by the scatterer. When the
scatterer is an axisymmetric thin flux tube, there are only three ways in which
energy can be extracted. (1) The single incident wave can excite tube waves which
subsequently propagate out of the acoustic cavity. This mechanism is a true absorption
process whereby acoustic energy is removed from the system. (2) The incident wave
can generate an acoustic jacket that surrounds the flux tube. This mechanism is only
significant for initial value problems and not for a spectral treatment as performed
here. (3) The incident wave is scattered into outgoing waves of a different mode order.
This last effect is mode mixing which is not true absorption as the acoustic energy
remains in the cavity, albeit at a different wavenumber.  We will assess the relative
importance of the first and third mechanisms and leave the second to subsequent work.

\subsection{Energy Budget of a Single Incident Wave}
\label{subsec:Ebudget}

Consider a single incident wave of order $n$, with a unit amplitude (${\cal A}_{mn} = 1$).
This incident wave generates a set of outgoing waves with mode orders $n'$. These
outgoing waves can be written as follows, where we have included both the far-field
scattered waves and the outgoing component of the incident wave,

\begin{equation}
    \Phi_{\rm out}(\omega;r,z) = \sum_{n'=0}^{\infty}
		\left[\frac{1}{2}\delta_{nn'}+S_0^{n \to n'}(\omega)\right] H_m^{(1)}(k_{n'}r) \; Q_{n'}(\omega;z) \; .
\end{equation}

\noindent The term in square brackets is the amplitude of the outgoing wave of mode
order $n'$, and is composed of a contribution from the incident wave ($\delta_{nn'}/2$)
and from a scattered wave ($S_0^{n\to n'}$). If the vertical eigenfunctions $Q_n$ are
normalized according to equation~\eqnref{eqn:orthog_discrete}, each outgoing wave carries
a lateral energy flux $F_{n'}$ with the following form,

\begin{equation}
    F_{n'} = \frac{1}{2} \, \rho_0 \, \omega^3 z_0  \left|\delta_{nn'}+2S_{0}^{n \to n'}\right|^2\; .
\end{equation}

\noindent Note, the energy flux only depends on the mode order $n'$ (and the azimuthal
order $m$) through the amplitude. Therefore, the fraction $\varepsilon_{nn'}$ of the
ingoing wave's energy that is re-emitted in mode order $n'$ has similar dependence on
the scattering matrix as exhibited by the absorption coefficient, 

\begin{equation} \label{eqn:Efrac}
	\varepsilon_{nn'} = \left|\delta_{nn'}+2S_0^{n \to n'}\right|^2 \; .
\end{equation}

For this example with only a single incident mode order, the absorption coefficient is
by definition 1 minus the diagonal elements of this fractional energy flux,

\begin{eqnarray}
	\alpha_n \equiv 1 - \varepsilon_{nn} &=& 1 - \left|1 + 2S_0^{n \to n}\right|^2
\label{eqn:alpha}
\\
	&=& -4{\rm Re}\left\{S_0^{n \to n}\right\} - 4 \left|S_0^{n \to n}\right|^2 \; .
\nonumber
\end{eqnarray}

\noindent Of course this equation is identical to equation~\eqnref{eqn:abscoef_singlemode}.
By energy conservation, we can decompose this absorption coefficient into two parts, a
true absorption from the excitation and escape of sausage waves $W_n$ and a redistribution
term arising from mode mixing $M_n$,

\begin{equation}
	\alpha_n = W_n + M_n .
\end{equation}

\noindent The contribution from mode mixing $M_n$ can be directly written down using
the off-diagonal elements of the fractional energy loss $\varepsilon_{nn^\prime}$,

\begin{equation} \label{eqn:Emodemix}
	M_n = \sum_{n^\prime \ne n} \varepsilon_{nn^\prime}
		= 4 \sum_{n^\prime \ne n} \left| S_0^{n\to n^\prime} \right|^2 .
\end{equation}

\noindent Therefore, by combining equations~\eqnref{eqn:alpha} and \eqnref{eqn:Emodemix},
the loss of energy due to the excitation of tube waves must have the following expression,

\begin{equation} \label{eqn:Etubewaves}
	W_n = -4{\rm Re}\left\{S_0^{n \to n}\right\} - 4 \sum_{n'=0}^\infty \left|S_0^{n \to n^\prime}\right|^2 .
\end{equation}

Before we can use any of these expressions, we need to consider the order of each of their
terms. Since we have computed the scattering matrix to only leading order in the radius of 
the tube, ${\rm O}\left(R^2\right)$, for self-consistency we must drop the last term in
equations~\eqnref{eqn:alpha} and \eqnref{eqn:Etubewaves},

\begin{eqnarray} \label{eqn:Efrac_OrderR2}
	\alpha_{n}(\omega) = - 4{\rm Re}\left\{S_0^{n \to n}\right\} + {\rm O}(R^4) \; ,
\\
	W_n(\omega) = -4 {\rm Re}\left\{S_0^{n \to n}\right\} + {\rm O}(R^4) \; .
\end{eqnarray}

\noindent To lowest order, the absorption coefficient and the excitation rate of tube waves
are identical. Therefore, our calculation of the absorption coefficient is only compatible
with tube-wave generation. The terms due to mode mixing appear at higher orders which are not
represented self-consistently. Fortunately, we can evaluate the mode mixing terms separately
using the far-field scattering matrix through equations~\eqnref{eqn:Efrac} and
\eqnref{eqn:Emodemix}.

Figure~\ref{fig:absorb} illustrates the absorption coefficient (solid curves) evaluated
through the use of equation~\eqnref{eqn:Efrac_OrderR2}. The fraction of the incident
energy, $M_n$, mixed
into all other modes is represented with the dotted curves. In Figure~\ref{fig:modemix},
the energy $\varepsilon_{nn^\prime}$ that is scattered into each individual mode is shown.
From this figure, one can see that the amount of energy lost to mode mixing decreases as
the order of either the scattered mode $n'$ or the incident mode $n$ increases. The symmetry
of the scattering matrix leads to the symmetry of $\varepsilon_{nn'}$, which is apparent
in Figure~\ref{fig:modemix}$a$ as the equal height of pairs of symbols of different color
and in Figure~\ref{fig:modemix}$b$ as the overlying of curves.

For low-frequencies, the mode-mixing losses are rather inconsequential compared to
the sausage-wave losses. However, for higher frequencies the amount of energy lost
to mode mixing can be a relative high fraction of the total energy loss, perhaps
reaching as high as a third for some parameters. Furthermore, for the $f$ mode at
high frequencies the excitation of sausage waves becomes anomolously inefficient
and the mode-mixing terms can be the dominant source of energy loss. A more in depth
discussion of tube wave excitation by the $f$ mode appears later in
\S\ref{subsec:absorption}.


\section{Phase Shifts and Phase Correlations}
\label{sec:phases}
\setcounter{equation}{0}

The computation of the scattering matrix also allows us to estimate phase shifts
$\Delta_n$ through equation~\eqnref{eqn:phaseshift_singlemode}. These phase
shifts are illustrated in Figure~\ref{fig:phase_shift}. For $p$ modes, both positive
and negative phase shifts are possible, although all shifts become positive for
sufficiently high frequency---which, depending on the value of $\beta$ may be beyond
the physically relevant band of frequencies. The $f$ mode is fundamentally different.
Not only is the phase shift for the $f$ mode negative for all frequencies, but the
phase shift is large, reaching values as high as 15$^\circ$ for high frequencies
and for tubes with low values of the plasma parameter $\beta$.

The phases of all the scattered waves are correlated in a precise manner with the
phase of the incident wave. The difference in phase $\Delta\phi_{nn'}$ between each
individual scattered wave and the incident wave can also be derived from the scattering
matrix,

\begin{equation} \label{eqn:phase_difference}
	\Delta\phi_{nn'} = \arg\left\{S_0^{n \to n'}\right\} \; .
\end{equation}

\noindent These relative wave phases are uniformly small (see Figure~\ref{fig:scatphase}).
The difference is positive for $n+n'$ even and negative for $n+n'$ odd. The symmetry
of the scattering matrix to exchange of $n$ and $n'$ appears in Figure~\ref{fig:scatphase}
as the overlying of curves and the equal height of symbols.
In general, scattering to and from the $f$ mode results in larger phase differences,
exceeding scattering between $p$ modes by an order of magnitude.

\section{Discussion}
\label{sec:discussion}
\setcounter{equation}{0}

We have calculated by semi-analytic means the scattering matrix for axisymmetric
acoustic waves encountering a vertically-aligned, magnetic tube. We have assumed
that the tube is slender and invoked the thin flux-tube approximation to compute
the scattering matrix to leading order in the tube's radius $R$
for mode orders up to $n = 10$. In the following two sections we discuss the
implications of the results of our calculation, both in terms of the energy budget
of the incident and scattered waves and in terms of the phases of the outgoing waves.

\subsection{Absorption and Energy}
\label{subsec:absorption}

When encountering a thin flux tube of the type discussed here, a $p$ mode of a
particular radial order $n$ may lose energy in either of two ways. The incident
acoustic wave can excite MHD tube waves on the magnetic fibril that freely propagate
along the tube carrying energy out of the acoustic cavity. This represents a true
absorption as acoustic energy is removed from the wavefield. The tube may also
scatter the incident $p$ mode into outgoing waves of different radial order $n'$.
Mode mixing of this sort does not represent a true absorption as the
energy  is merely redistributed between radial orders. Despite this fact, mode mixing
can modify the observed absorption coefficient, as the definition of this coefficient
does not take power redistribution into account. These two mechanisms are both
inherently weak for thin tubes. The thin tube approximation mandates that the
excitation of tube waves generates an absorption coefficient that scales linearly
with the magnitude of the scattering matrix, which in turn scales like the square
of the tube's radius $R^2$, a small quantity by definition. The energy losses due
to mode mixing are even smaller, scaling as the square of the magnitude of the
scattering matrix, or as $R^4$. Therefore, mode mixing can only be important
energetically when other sources of absorption (i.e., tube wave excitation) are
peculiarly small. Over the physically relevant range of frequencies ($< 6$ mHz),
our calculations reveal that mode-mixing dominates the absorption coefficient
only for high-frequency $f$ modes interacting with low-$\beta$ tubes.

Any model with a reflective upper boundary can have special frequencies for which
the excitation of tube waves is unusually tiny. If the upper boundary reflects all
wave energy, the excitation can formally be zero at these frequencies. Such ``excitation
nulls" arise because of total destructive interference. A localized driver, such
as an incident $p$ mode, excites waves that propagate both up and down the tube
with equal energy. At these special frequencies, the upward propagating wave reflects
off the upper surface with just the phase needed to cancel the downward propagating
wave \citep{Jain:2009, Crouch:1999}. In such a circumstance the downward energy
flux vanishes and no energy is carried away from the acoustic cavity by tube waves.
Near such an excitation null, the energy losses arising from mode mixing could be the
dominant source of energy loss---not because mode mixing is large, but because the
other source of absorption is small.

For the stress-free upper boundary condition applied here and for reasonable atmospheric
and tube parameters, excitation nulls do not appear within the physically relevant
frequency band ($\omega/2\pi < 6$ mHz). However, formally for the $f$ mode such a
null exists in the limit of infinite frequency and it has extremely broad influence,
causing the absorption coefficient to asymptote to zero as the frequency becomes large.
The frequency regime for which this asymptotic behavior is valid depends on $\beta$,
with lower $\beta$ pushing the behavior to lower frequencies. Physically, such behavior
exists because the $f$ mode is a surface wave that becomes increasingly confined to
the photospheric layers as the frequency increases. The $f$ mode's evanescence length
$k^{-1}$ can be derived directly from its dispersion relation $\omega^2 = gk$, revealing
that the vertical extent of the $f$ mode scales as $\omega^{-2}$. The wavelength $L$
of the sausage wave at the photospheric surface can be derived from
Equation~\eqnref{eqn:HomogeneousTubeWave},

\begin{equation}
	L = \left[\frac{(\gamma-1) g}{\left(1 + \gamma\beta/2\right) z_0}\right]^{1/2} \frac{\pi}{\omega} \; .
\end{equation}

\noindent As the frequency increases, $L$ shrinks, scaling as $\omega^{-1}$; however,
it decreases more slowly than the scale length of the $f$ mode decreases (which scales
as $\omega^{-2}$). Therefore, as the frequency increases, compared to the tube wave
the driver looks more and more like a delta function located at the surface. A
delta-function driver generates waves that propagate in each direction with identical
amplitude and phase. If one of those components subsequently reflects off a surface
and passes back through the driving layer, there is the potential for destructive
interference (as discussed previously). Destructive interference occurs if the phase
of the reflected wave advances by an odd integer factor of $\pi$. This phase advance
can come from a phase change on reflection or from propagation of the wave to the
reflecting layer and back. For the $f$ mode, one can show that the phase change on
reflection asymptotes to $-\pi$ as the frequency becomes large
(see Appendix~\ref{app:HighFrequency}). Therefore, the $f$ mode, in the limit of infinite
frequency, satisfies the necessary resonance condition because the distance between
the driving layer and the photospheric reflection layer vanishes.

The $p$ modes do not evince such high-frequency behavior for the simple reason that
at high frequency, the depth of the cavity becomes roughly constant with frequency.
The wavenumber eigenvalue $k_n(\omega)$ scales linearly with the frequency, thereby
causing the horizontal phase speed $\omega/k$ and the lower turning point of the $p$
mode to approach constants.
Since, the depth of the cavity and hence the vertical extent of the driving region
does not change as the frequency changes, the wavelength of the sausage waves becomes
much shorter than that of the $p$ modes. Thus, for incident $p$ modes, the driving
region remains vertically distributed and the arguments that we made previously for the
$f$ mode are not valid.

Figure~\ref{fig:highfreq} illustrates this
behavior for $\beta = 0.1$. While the figure extends to frequencies as large as
30 mHz, one must bear in mind that frequencies greater than 6 mHz are greater than
the sun's photospheric acoustic cutoff; thus, such waves are no longer trapped and
the construction of our solar model is suspect. Further, one must be careful to not
over-interpret  the results at such extreme frequencies as the perturbation scheme has
become invalid (see Figure~\ref{fig:highfreq}). Despite this, for the $f$ mode,
trapped waves of low frequency (where the perturbation analysis remains valid) can
still sense the presence of the excitation null at infinity, causing a
high-frequency fall-off of the absorption. The frequency of waves
affected depends on the plasma parameter $\beta$ of the tube because the wavelength
of the tube wave depends on $\beta$. We can estimate the lowest frequencies that
should be affected by requiring that the wavelength of the sausage waves $L$ be
two or three times larger than $f$ mode's evanescence length $k^{-1}$.
If we require that the ratio of these length scales satisies $kL > 2$, we discern
that for $\beta = 0.1$ we would expect the high frequency regime to exist for frequencies
above 3 mHz, whereas for $\beta = 10$ the regime begins around 10 mHz.

This discussion only holds for absorption and mode mixing resulting from the excitation
of the slow sausage wave. Similar calculations performed with fast kink oscillations
reveal that excitation nulls can appear for all incident mode orders at frequencies
less than 6 mHz \citep{Jain:2009, Crouch:1999}; also see \cite{Tirry:2000} for a similar
phenomenon in an unstratified model. Further, the calculations of the scattering matrix
elements arising from coupling to kink 
oscillations ($S_{\pm 1}^{n\to n'}$) \citep{Hanasoge:2008} show significant oscillatory
modulation in the magnitudes as a function of frequency, whereas our scattering calculation
for the axisymmetric sausage wave ($S_0^{n \to n'}$) show smooth frequency variation. In
fact one could fit the magnitude of the scattering matrix calculated here with a single
power law in frequency that would be roughly valid over all but the highest frequencies.

In a real plage, we do not expect such excitation nulls to have an easily identifiable
effect other than a diminution of absorption at high frequencies for the $f$ mode. Not
only does mode mixing tend to fill in the absorption profile, but the plage is likely to
be comprised of a large number of tubes with a distribution of properties, including
photospheric radius $R_0$ and plasma parameter $\beta$. Therefore, each tube in the
collection will have suppressed absorption over a different range of frequencies. The
added affect of all of the tubes will be to mitigate the effect of nulls \citep{Jain:2011}.
In general, we expect most flux tubes in a plage to have a near equipartition of magnetic
and gas pressure, thus tubes with $\beta \approx 1.0$ should predominate. A quick examination
of Figure~\ref{fig:absorb} reveals that only the most extreme frequencies are likely to
be affected.

To summarize, we find that for thin tubes, mode mixing is a small effect compared to
tube wave excitation. The scattering matrix itself has elements that are proportional
to the cross-sectional area of the tube. Tube wave excitation extracts energy at a rate
proportional to the scattering matrix (or proportional to the cross-sectional area of
the tube) while mode mixing redistributes energy at a rate proportional to the square
of the scattering matrix.

\subsection{Wave Phases}
\label{subsec:phases}

Phase shifts are generally defined as the difference in phase between the outgoing wave
in the presence of the scatterer and the phase that would have existed in the absence of
the scatterer. A magnetic tube can generate a phase shift by three distinct mechanisms:
(1) The ingoing acoustic wave reflects off the lateral surface of the tube, shortening
the total path length traveled by the wave. (In the absence of the tube, the wave would
travel all the way to the coordinate axis and back.) Since the wave travels a shorter
distance, the presence of the tube causes a phase lag resulting in a negative phase shift
or travel-time difference. (2) The wave is at least partially transmitted through the
lateral surface and propagates at a different speed within the magnetic tube. If the
phase speed increases, a negative phase shift is achieved. (3) Upon reflection off the
lateral surface of the tube, the wave can undergo a phase change.  Off course if a
flux-tube boundary could be treated as a rigid, impenetrable surface, the phase change
would be 180 degrees. But, in general a flux tube is elastic and the phase change may
differ from 180 degrees and may be either positive or negative.

For our thin flux-tube model, the only wave permitted within the tube is a sausage wave
corresponding to a slow surface wave \citep[see][]{Roberts:1978, Edwin:1983}. For
such a wave, the thin flux-tube approximation precludes lateral variation by implicitly
assuming that the wave crossing time is much shorter than the wave period. Thus, for
all tubes the crossing time for the tube should be ignored, resulting in a common negative
phase shift, independent of $\beta$. Obviously the phase shift whould be equal to the phase
lag that would have occured in the unmagnetized model for a wave propagating a distance
equal to the tube diameter. This shows that for thin flux tubes, mechanisms 1 and 2 discussed
in the previous paragraph are essentially the same. This is not true for thicker tubes for
which the travel time across the tube is significant. A rough estimate of the expected phase
shift would be $-2kR_0$, where $R_0$ is the photospheric radius of the tube and we have
{\sl assumed} that the phase lag should be evaluated
at the photospheric surface. This is clearly appropriate for the $f$ mode as it is a surface
wave confined to this layer. For $p$ modes one should probably consider a depth weighted
average of the tube radius. However, for our purposes, the previous expression will suffice.
Estimates made this way for the $f$ mode give a phase shift of $\Delta_0 = -2\omega^2 R_0/g$,
which for waves at 3 mHz give a phase shift of -15$^\circ$. This surprisingly large shift
is about a factor of four larger than the phase shifts we obtain with our model (see
\ref{fig:phase_shift}), indicating that a phase change upon reflection is likely to be at
work as well. We point
out that despite the large size of the estimated shift, it is quite similar in magnitude
and of the same sign as actually measured phase shifts for small magnetic features.
\cite{Duvall:2006} measured the average scattering from roughly 2500 small magnetic elements, and
generated a typical travel time kernel from their measurements. One can obtain a typical
phase shift of -14$^\circ$ by the following process. Multiply the largest value of the measured
kernel in Figure 6 from \cite{Duvall:2006} (roughly -20 s kG$^{-1}$ Mm$^{-1}$) by the average
magnetic flux of the studied features (0.6 kG Mm$^2$) to obtain a travel time lag of -12 s.
Convert this time lag into a phase shift by dividing by a wave period (300 s) and multiplying
by 360 degrees.

The phase change that occurs at reflection from the lateral boundary of the flux tube is
difficult to estimate, as the reflection depends on both the incident wave and the driven
sausage wave over a range of heights. We would like to receive guidance from calculations
performed for unstratified, thin tubes, for which analytic solutions are possible
\citep[e.g.,][]{Edwin:1983}. However, the regime in frequency--wavenumber space for which
slow surface waves exist and for which propagating external acoustic waves exist are disjoint.
Thus, for unstratified tubes a propagating acoustic wave cannot excite slow surface waves.
The stratification in our model is what makes the coupling possible. We can safely say,
however, that the phase shift caused by mechanisms 1 and 2 always result in negative shifts
for thin tubes. Further, we would expect the phase change upon reflection (mechanism 3)
to increase for tubes of higher $\beta$ since such tubes are more elastic. For the
$f$ mode this is indeed what we achieve; the phase shift is large and negative at all
frequencies (probably as a result of a reduced path length) and becomes less negative
as $\beta$ increases (probably due to an increasing positive phase change at reflection).
However, for $p$ modes we obtain shifts of both signs; negative phase shifts at low
frequency and positive shifts at high frequency. Clearly for the $p$ modes a positive
phase change upon reflection at the flux-tube boundary must occur for at least high
frequencies.


\acknowledgments

We thank Tom Duvall for his helpful suggestions in how to convert measured travel-time
kernels into phase-shift estimates. We thank Tom Bogdan for a useful conversation
during which the proper ordering of the different wave components became apparent.
BWH acknowledges support from NASA through grants NNG05GM83G,
NNX08AJ08G, NNX08AQ28G and NNX09AB04G. RJ acknowledges STFC (UK) for partial funding.

\appendix

\section{The Field-Free Atmosphere and its Eigenfunctions}
\label{app:eigenfunctions}
\setcounter{equation}{0}

The field-free atmosphere that surrounds the magnetic flux tube is treated as a plane-parallel,
neutrally stable polytrope. This atmosphere has been utilized extensively in the past
as its a good representation of a stellar convection zone and simplifies the wavefield
by imposing a condition of irrotationality. Gravity is uniform, acting in the downwards
direction $\bvec{g} = -g\hat{z}$, with the height coordinate $z$ increasing upward. The
atmosphere is polytropic below the height $z=-z_0$, which corresponds to the model's
photosphere. The gas pressure, mass density and sound speed vary with height as power
laws. If we employ a nondimensional depth $s = -z/z_0$, the background thermodynamic
variables become,

\begin{eqnarray*}
    P_\e &=& \frac{g z_0 \rho_0}{a+1} \; s^{a+1}
                = P_0 \; s^{a+1} \; ,    \\
                                                           	\\
    \rho_\e &=& \rho_0 \; s^a \; ,	\\
								\\
    c_\e^2 &=& \frac{gz_0}{a}\; s \;.
\end{eqnarray*}

The quantities $\rho_0$ and $P_0$ are the photospheric values of the mass density and
gas pressure. The value of the polytropic index $a$ is set such that the stratification
is neutrally stable to convection; this requires $a = 1/(\gamma-1)$, where $\gamma$ is
the ratio of specific heats. Above the photosphere $z = -z_0$, we assume the existence
of a hot vacuum ($\rho_\e \to 0$ with $T_\e \to \infty$), with the property that the gas
pressure ($P_\e \sim \rho_\e T_\e$) is finite and continuous across the $z = -z_0$ layer.

Following \cite{Bogdan:1996} and \cite{Hindman:2008}, we specify the depth of the
photosphere $z_0$ and
the photospheric density $\rho_0$ (and therefore the reference pressure $P_0$)
by matching our model photosphere to the $\tau_{5000} = 1$ level of a solar model
by \cite{Maltby:1986}. At this layer in the solar model $g=2.775 \times 10^4$
cm s$^{-2}$, $\rho_0 = 2.78 \times 10^{-7}$ g cm$^{-3}$, and $P_0 = 1.21 \times 10^5$
g cm$^{-1}$ s$^{-2}$. We adopt a polytropic index of $a = 1.5$, which is consistent
with an adiabatic index of $\gamma = 5/3$.

Since the buoyancy frequency in such an atmosphere is zero by definition, acoustic
waves propagating within the atmosphere are irrotational. This permits such waves to be
expressed using a displacement potential $\Phi$. If we assume oscillatory time dependence
$~e^{-i\omega t}$ and harmonic functions with wavenumber $k$ in the horizontal direction,
the vertical portion of the acoustic eigenfunction satisfies the following Sturm-Liouville
equation,

\begin{eqnarray}
    	\left[ \frac{d}{ds}\left(s^a \frac{d}{ds}\right) + \nu^2 s^{a-1} -\frac{\lambda^2}{4} s^a\right]Q(\nu;s) &=& 0 \; ,
	\label{eqn:sturm_liouville}	\\
	\lambda &\equiv& 2 k z_0 \; ,		\\
	\nu^2 &\equiv& \frac{a \omega^2 z_0}{g} \; ,
\end{eqnarray}

\noindent where $\lambda$ is a dimensionless horizontal wavenumber, $\nu$ is a dimensionless
frequency and $Q$ represents either a discrete eigenfunction $Q_n(\nu;s)$ for real
$\lambda = \lambda_n = 2 k_n z_0$ or a spectrum of jacket modes $q(\nu,\Lambda;s)$ for
continuous, imaginary values of $\lambda = -2i \Lambda z_0$.

The general solutions to this equation are proportional to Whittaker $M$ and $W$ functions
\citep{Whittaker:1952, Abramowitz:1964}. The discrete $f$ and $p$ modes reside within an
acoustic cavity and therefore become evanescent deep in the atmosphere. Thus, the $W$
functions are relevant as they decay exponentially as $s\to\infty$,

\begin{equation}
            Q_n(\nu;s) = C_n(\nu) \; s^{-(\mu+1/2)} \; W_{\kappa_{n},\mu}(\lambda_n s) \; .
\end{equation}

\noindent Here $C_n(\nu)$ is a normalization constant and

\begin{displaymath}
    \mu \equiv (a-1)/2\; , \qquad \kappa_n \equiv {\displaystyle \frac{\nu^2}{\lambda_n}} \; .
\end{displaymath}

\noindent The quantization of the horizontal wavenumber $k_n$ (or equivalently $\lambda_n$)
arises from the requirement that the Lagrangian pressure perturbation vanish at the model
photosphere.

The laterally evanescent jacket waves are surface waves (with a lateral decay length
of $\Lambda^{-1}$) that propagate up and down the outside of the flux tube. They are not
trapped within the acoustic cavity and can freely propagate to any depth. If we select
only those jacket waves that satisfy the condition that the Lagrangian pressure perturbation
vanishes at the model photosphere, the jacket waves are standing waves that can be
represented as a sum of Whittaker $M$ functions with imaginary argument
\citep[see][]{Bogdan:1995},

\begin{equation} \label{eqn:jacketmodes}
       q(\nu,\Lambda;s) = C(\nu,\lambda) \; s^{-(\mu+1/2)} \;
            \left[\Gamma_{\kappa,\mu}(\lambda) \; M_{\kappa,\mu}(\lambda s)
            + M_{\kappa,-\mu}(\lambda s)\right] \; ,
\end{equation}

\noindent where $C(\nu,\lambda)$ is a normalization constant and

\begin{displaymath}
    \kappa \equiv \frac{\nu^2}{\lambda} \; , \qquad
        \Gamma_{\kappa,\mu}(\lambda) \equiv \frac{16(1+\mu)(1+2\mu)^2}{(1+2\mu)^2-\kappa^2}
            \frac{M_{\kappa,-(1+\mu)}(\lambda)}{M_{\kappa,1+\mu}(\lambda)} \; .
\end{displaymath}

\noindent Unlike the $f$ and $p$ modes, for which the dimensionless horizontal wavenumber
$\lambda$ is purely real and only takes on discrete values $\lambda_n$, for the jacket modes
$\lambda$ is purely imaginary and continuous. Therefore, the argument and $\kappa$-index
of the Whittaker functions in equation~\eqnref{eqn:jacketmodes} are purely imaginary as well.

Since both the discrete modes and the jacket modes satisfy Hermitian boundary conditions
(vanishing Lagrangian pressure perturbation at both the photosphere $s=1$ and at infinity
$s\to\infty$), we automatically know that the discrete, acoustic modes and the jacket modes
are all mutually orthogonal with a weight function $s^a$ (which is proportional to the mass
density),

\begin{eqnarray}
    \int_1^\infty ds \; s^a \; Q_n(\nu; s) \; Q_{n'}(\nu; s) &=& \delta_{nn'}  \; ,
        \label{eqn:orthog_discrete}                         \\
    \int_1^\infty ds \; s^a \; Q_n(\nu; s) \; q(\nu,\Lambda; s) &=& 0  \; ,
        \label{eqn:orthog_cross}                           \\
    \int_1^\infty ds \; s^a \; q^*(\nu,\Lambda; s) \; q(\nu,\Lambda'; s) &=&
            \delta(\Lambda-\Lambda') \; .
        \label{eqn:orthog_jacket}
\end{eqnarray}

\noindent In writing these last equations, we have chosen $C_n(\nu)$ and $C(\nu,\lambda)$
such that the discrete modes and jacket modes are orthonormal.

\section{Sausage Waves}
\label{app:sausagewaves}
\setcounter{equation}{0}

We have assumed the magnetic fibril is untwisted, straight, axisymmetric, vertically
aligned and thin. Thin flux tubes are unable to support internal lateral structure
and hence the total pressure is uniform across the tube and continuous with its
external value. If we further assume that the tube is the same temperature as its
surroundings, the total pressure must have the same scale height inside and outside
the tube and the plasma parameter $\beta$, defined as the ratio of the gas pressure
to the magnetic pressure, is constant with height inside the tube. Since we only
consider tubes below the photosphere (i.e., $z < -z_0$), we need not worry about
the rapid flaring of the tubes into a magnetic canopy that occurs within the
chromosphere.

We ignore lateral variation of the magnetic field strength, and describe the tube's
internal gas pressure $P(z)$, mass density $\rho(z)$, sound speed $c(z)$, and field
strength $B(z)$ by their axial values. These four quantities as well as the tube's
radius $R(z)$ can be described uniquely by the total magnetic flux $\Theta$ contained
by the tube and by the plasma $\beta$,

\begin{eqnarray}
    P &=& \frac{\beta}{\beta+1} P_\e \; ,             		\\
\nonumber							\\
    \rho &=& \frac{\beta}{\beta+1} \rho_\e \; ,         	\\
\nonumber							\\
    c &=& c_\e \; ,                                     	\\
\nonumber							\\
    \frac{B^2}{8\pi} &=& \frac{1}{\beta+1} P_\e \;,   		\\
\nonumber							\\
    \pi R^2 &=& {\displaystyle \frac{\Theta}{B}} =
                \left(\frac{\beta+1}{8\pi P_\e}\right)^{1/2} \Theta \; .
          \label{eqn:cross-area}
\end{eqnarray}

Observations indicate that typical field strengths for small photospheric flux tubes
are between 1 and 2 kG within both quiet sun and plage
\citep[e.g.,][]{Lagg:2010, Martinez-Pillet:1997}. Such field strengths indicate rough
equipartition between magnetic and gas pressure. For a flux tube with $\beta=1$ embedded
in the polytropic atmosphere described in  Appendix~\ref{app:eigenfunctions}, the
magnetic field strength at the photosphere is $B_0 = 1.2$ kG. By inserting the external
pressure $P_{\rm ext}$ into the expression for the cross-sectional area of the
tube~\eqnref{eqn:cross-area}, one obtains the explicit dependence of the tube radius
on depth,

\begin{equation} \label{eqn:TubeRadius}
    R(s) = R_0 s^{-(a+1)/4}
\end{equation}

\noindent where $R_0 = \left(\Theta/\pi B_0\right)^{1/2}$ is the photospheric
radius of the tube. For all illustrations we
will assume $R_0 = 100$ km, a fairly typical value for the size of
magnetic elements within solar plage \citep{Lagg:2010}.

Thin flux tubes support a longitudinal magnetoacoustic wave known as the sausage wave.
Here, these oscillations are driven by the pressure perturbation imposed by external
$f$- and $p$-mode oscillations \citep[e.g., ][]{Roberts:1978}. Using the formulation of
\cite{Hindman:2008}, to lowest order in the radius of the flux tube, the vertical displacement $\zeta_\parallel$
within the tube can be described by the following equation,

\begin{equation} \label{eqn:parallel}
    \left\{ \dparn{t}{2} - c_T^2\dparn{z}{2} +
        \frac{\gamma g}{2}\frac{c_T^2}{c^2}\dpar{z} \right\} \zeta_\parallel =
        \frac{\rho_\e}{\rho} \frac{c_T^2}{V_{\rm A}^2}
            \left. \frac{\partial^3\Phi_{\rm inc}}{\partial z\partial t^2} \right|_{r=0} \;,
\end{equation}

\noindent where $c_T$ is the cusp or tube speed which depends on both the sound
speed $c$ and Alfv\'en speed $V_{\rm A}$ within the tube,

\begin{displaymath}
    c_T^2 = \frac{c^2 V_{\rm A}^2}{c^2 + V_{\rm A}^2} \; ,  \qquad
    V_{\rm A}^2 = \frac{B^2}{4\pi \rho} \; .
\end{displaymath}

\noindent The right-hand side of Equation~\eqnref{eqn:parallel}
is due to the forcing by the incident wave and is evaluated at the flux tube's axis,
$r=0$. This forcing arises from the enforcement of continuity of total pressure
across the boundary of the flux tube.  We are allowed to consider only the incident
waves in this pressure balance because the far-field and near-field scattering amplitudes
are smaller than the incident wave by a factor of $R^2 \ln R$.

The forcing provided by each individual incident $p$ mode can be obtained by evaluating
the expression for the incident wavefield, Equation~\eqnref{eqn:incident}, at $r = 0$
and inserting the result into the sausage wave equation. After Fourier transforming in
time $t$, one obtains

\begin{equation} \label{eqn:sausage}
    \left\{ c_T^2\dparn{z}{2} - \frac{\gamma g}{2}\frac{c_T^2}{c^2}\dpar{z} + \omega^2 \right\} \zeta_\parallel =
        \frac{\rho_\e}{\rho} \frac{c_T^2}{V_{\rm A}^2} \omega^2
        \sum_{n=0}^\infty \; {\cal A}_{n}(\omega) \frac{dQ_n(\omega;z)}{dz} \; .
\end{equation}

\noindent Note that only the $p$ modes with $m = 0$ contribute to the forcing of the
sausage waves. All other azimuthal components vanish in the limit $r \to 0$.

We adopt a non-dimensional form for this equation by the following substitutions,

\begin{displaymath}
    s = -\frac{z}{z_0} \; , \qquad
    \epsilon = \frac{2+\gamma\beta}{2} \; , \qquad
    f_n(s) = -\frac{\gamma(\beta+1)}{2} \frac{\nu^2}{s}\frac{dQ_n(\nu;s)}{ds} \,
\end{displaymath}

\noindent producing

\begin{equation}
    \left(\frac{d^2}{ds^2} + \frac{\mu+1}{s}\frac{d}{ds} + \frac{\nu^2 \epsilon}{s}\right)\zeta_\parallel(s) =
\sum_{n=0}^\infty\frac{{\cal A}_{n}}{z_0}f_n(s) \; .
\end{equation}

\noindent The solution of this equation is expressed using a Green's function,

\begin{eqnarray}
    \zeta_\parallel(s) &=& \sum_{n=0}^\infty \frac{{\cal A}_{n}}{z_0^2} \zeta_{\parallel,n}(s) \; ,
        \label{eqn:zetacomponents}
\\
    \zeta_{\parallel,n}(s) &=& -\frac{i\pi}{2} z_0
            \left\{\psi_\parallel(s)\left[\Omega_n+{\cal J}_n^*(s)\right]
                + \psi_\parallel^*(s)\left[{\cal I}_n-{\cal J}_n(s)\right]
            \right\} \; .
	\label{eqn:zetan}
\end{eqnarray}

\noindent In this equation, $\psi_\parallel(s)$ and $\psi_\parallel^*(s)$ are the two
solutions to the homogeneous equation, ${\cal J}_n(s)$ and ${\cal I}_n$ are interaction
integrals over the driver, and $\Omega_n$ is a parameter that specifies the complex
amplitude of the wave that reflects from the upper surface and thus determines
the boundary condition at the model photosphere \citep[see][for details]{Hindman:2008},

\begin{eqnarray}
    \psi_\parallel(s) &=& s^{-\mu/2}H_\mu^{(1)}(2\nu\sqrt{\epsilon s}) \; ,
	\label{eqn:HomogeneousTubeWave}
\\
    {\cal J}_n(s) &\equiv& \int_1^s d\sigma\; \sigma^{\mu+1} \psi_\parallel(\sigma) f_n(\sigma) \; ,
\\
    {\cal I}_n \equiv \lim_{s\to\infty}  {\cal J}_n(s) &=&
        \int_1^\infty d\sigma \; \sigma^{\mu+1} \psi_\parallel(\sigma) f_n(\sigma) \; .
	\label{eqn:InteractionIntegral}
\end{eqnarray}

If we apply a stress-free boundary condition on the sausage wave at the photosphere,
the boundary condition parameter has the following form,

\begin{eqnarray}
    \Omega_n &=& i\frac{\gamma(\beta+1)}{\pi} \; \nu^2 \; \frac{{\cal Q}_n}{\cal H}
        - \frac{{\cal H}^*}{\cal H}{\cal I}_n \; ,    \\
    {\cal H} &\equiv&  \nu\sqrt{\epsilon}H_{\mu+1}^{(1)}(2\nu\sqrt{\epsilon})
        + (\beta+1)(\mu+1)H_\mu^{(1)}(2\nu\sqrt{\epsilon}) \; , \\
    {\cal Q}_n &\equiv& \left.Q_n(\nu;s)\right|_{s=1} \; .
\end{eqnarray}

\noindent Figure~\ref{fig:wave_functions}$a,b$ illustrates the resulting
sausage wave as well as the driving $p$-mode eigenfunction when the incident
wave is a $p_2$ wave.

We can derive an expression for the normal displacement inside the tube using the definition
for the divergence of the displacement vector $\bvec{\xi}$ in cylindrical coordinates,

\begin{equation}
    \nabla\cdot\bvec{\xi_\parallel} = \frac{\partial h_\parallel}{\partial r}
        + \frac{h_\parallel}{r} + \frac{\partial \zeta_\parallel}{\partial z} \; ,          \\
\end{equation}

\noindent where $h_\parallel$ is the radial displacement. In this equation we
have utilized the axisymmetry of the sausage mode. This can be combined with the
Taylor series expansion of the radial displacement,

\begin{equation} 
    h_\parallel(\omega;r,z) = r \left.\frac{\partial h_\parallel(\omega;r,z)}{\partial r}\right|_{r=0} + \cdots,
\end{equation}

\noindent which we evaluate at $r=R(z)$ to obtain

\begin{equation}\label{eqn:hpar}
    h_\parallel(\omega;R,z) = \frac{R(z)}{2} \left(\nabla\cdot\bvec{\xi_\parallel}
        - \frac{\partial \zeta_\parallel}{\partial z}\right) \; .
\end{equation}

\noindent If we now use Equation A9 from \cite{Bogdan:1996}, which is a statement about pressure
equilibration of the tube with its external environment,

\begin{equation}
    \nabla\cdot\bvec{\xi_\parallel}=\frac{\beta}{2+\gamma\beta}\left(-\frac{\delta P_{\rm inc}}{P}
        + \frac{2}{\beta}\frac{\partial \zeta_\parallel}{\partial z} + \frac{g\rho_\e}{P}\zeta_\parallel\right).
\end{equation}

\noindent and combine equations~\eqnref{eqn:Ncomp} and \eqnref{eqn:hpar}, we obtain after
some manipulation

\begin{equation}
    N_\parallel(\nu;s) = \frac{R(s)}{4} \; \frac{\gamma \beta}{\epsilon z_0}
        \left[\frac{d \zeta_\parallel}{ds} + \frac{\mu}{s}\zeta_\parallel(s)
        - \frac{\beta+1}{\beta}\frac{\nu^2}{sz_0} \Phi_{\rm inc}(\nu;s)\right] .
\end{equation}

This displacement can be decomposed into seperate components driven by each incident mode,
by utilizing Equations~\eqnref{eqn:incident} and \eqnref{eqn:zetacomponents},

\begin{eqnarray}
    N_\parallel(\nu;s) &=& \sum_{n=0}^\infty \frac{ {\cal A}_{n}}{z_0^2} \; N_{\parallel,n}(\nu;s) , \\
    N_{\parallel,n}(\nu;s) &=& \frac{R(s)}{4} \; \frac{\gamma \beta}{\epsilon z_0}
        \left[\frac{d \zeta_{\parallel,n}}{ds} + \frac{\mu}{s}\zeta_{\parallel,n}(s)
                - \frac{\beta+1}{\beta}\frac{\nu^2}{s} z_0 Q_n(\nu;s)\right] .
	\label{eqn:Nparn}
\end{eqnarray}

\section{High-Frequency Asymptotics for Incident $f$ Modes}
\label{app:HighFrequency}
\setcounter{equation}{0}

The eigenfunction for the $f$ mode is simple enough that we can perform high-frequency
asymptotics to derive approximate analytic solutions for the interaction integral
${\cal I}_0$ and all quantities derived from it.  The $f$ mode eigenfrequency is
always $\omega^2 = gk$ independent of the details of the stratification. The associated
eigenfunction is just an exponential that grows with height $Q_0(\omega;z) = \exp(k z)$.
In our dimensionless variables, these relations take the form $\nu^2 = a \lambda/2$ and
$Q_0(\nu;s) = \exp(-\nu^2 s/a)$. The $f$ mode's evanescence length is given by
$a/\nu^2$ and as the frequency becomes large, the evanescence length becomes
tiny, scaling as $\sim \nu^{-2}$. The $f$ mode's eigenfunction is, therefore, progressively
more confined to the photospheric surface as the frequency becomes large. The tube
wave on the other hand has a wavelength that diminishes much more slowly. From
equation~\eqnref{eqn:HomogeneousTubeWave} the wavelength of the sausage wave can
be seen to scale as $\sim \nu^{-1}$. Thus, in the interaction integral ${\cal I}_0$
---see equation~\eqnref{eqn:InteractionIntegral}, the integrand is only significant
near the lower limit of integration (near the photosphere) and the sausage wave
can be considered long wavelength.

\subsection{The Interaction Integral}
\label{subsec:I0}

After inserting the $f$-mode eigenfunction and the homogeneous sausage wave
solution---equation~\eqnref{eqn:HomogeneousTubeWave}---into
equation~\eqnref{eqn:InteractionIntegral}, we obtain

\begin{eqnarray}
	{\cal I}_0 &=& \frac{\gamma(\beta+1)}{2a} \nu^4 \int_1^\infty ds \; F(s) \, e^{-\nu^2 s/a} \; ,
	\label{eqn:I0}
\\
	F(s) &\equiv& s^{\mu/2} H_\mu^{(1)}\left(2\nu\varepsilon^{1/2} s^{1/2}\right) \; .
\end{eqnarray}

\noindent Since the exponential term in the integrand has significant magnitude
only very close to the lower limit of integration and the sausage wave is long
wavelength compared to the exponential scale length, we may expand the function
$F(s)$ in a Taylor series about $s = 1$,

\begin{equation}
	F(s) = F(1) + F'(1)(s-1) + \frac{F''(1)}{2} (s-1)^2 + \cdots
\end{equation}

\noindent If we define

\begin{equation} \label{eqn:little_h}
	h_\mu \equiv H_\mu^{(1)}\left(2\nu\varepsilon^{1/2}\right) \; ,
\end{equation}

\noindent and perform the necessary derivatives, the following expression is
obtained,

\begin{eqnarray}
	F(s) 	&=& h_\mu + \left(\mu h_\mu - \nu \varepsilon^{1/2} h_{\mu+1}\right) (s-1)
	\nonumber
\\
		& & +\frac{1}{2} \left[ \left( \frac{3}{2}\mu^2 - \mu - \nu^2\varepsilon \right) h_\mu
			-(\mu-1)\nu\varepsilon^{1/2} h_{\mu+1}\right] (s-1)^2 + \cdots
\end{eqnarray}

After inserting this expansion into the integral appearing in equation~\eqnref{eqn:I0},
each term in the expansion can be integrated separately by analytic means, producing
the following expansion for the interaction integral,

\begin{equation} \label{eqn:I0_asymp}
	{\cal I}_0 = \frac{\gamma(\beta+1)}{2} \nu^2 e^{-\nu^2/a} \left[h_\mu
			+ \frac{a}{\nu^2} \left(\mu h_\mu - \nu \varepsilon^{1/2} h_{\mu+1}\right)
			- \frac{a^2 \varepsilon}{\nu^2} h_\mu + {\rm O}\left(\nu^{-7/2}\right)\right] \; .
\end{equation}

\noindent The order of the missing terms was deduced by noting that for large frequency
the large-argument expansion of the Hankel function in equation~\eqnref{eqn:little_h}
is appropriate. Therefore, $h_\mu \sim \nu^{-1/2}$ and the leading order frequency
dependence for ${\cal I}_0$ is ${\rm O}(\nu^{3/2}e^{-\nu^2/a})$. To lowest order, the
interaction integral should decay rapidly as the frequency increases.

\subsection{Amplitude of the Downward Propagating Sausage Wave}
\label{subsec:DownwardAmp}

The complex amplitude of the wave that reflects from the photosphere is given by
the boundary condition parameter $\Omega$. For the stress-free upper boundary with
an $f$-mode incident wave, the boundary condition parameter $\Omega_0$ takes on the
following form,

\begin{equation} \label{eqn:Omega0}
	\Omega_0 = i\frac{\gamma(\beta+1)}{\pi} \; \nu^2 \frac{e^{-\nu^2/a}}{\cal H} - \frac{{\cal H}^*}{\cal H} {\cal I}_0 \; .
\end{equation}

\noindent If we now combine equations~\eqnref{eqn:I0_asymp} and \eqnref{eqn:Omega0}
we find asymptotically,

\begin{eqnarray}
	\Omega_0 = \frac{\gamma(\beta+1)}{2} \; \nu^2 \frac{e^{-\nu^2/2}}{\cal H}
		&& \left[ \frac{2i}{\pi} - \nu \varepsilon^{1/2} h_{\mu+1}^* h_\mu - (\mu+1)(\beta+1)h_\mu^* h_\mu \right.
	\nonumber
\\
		&& \qquad + a(\mu+1)(\beta+1) \nu^{-1} \varepsilon^{1/2} h_\mu^* h_{\mu+1} - a\mu\nu^{-1}\varepsilon^{1/2} h_{\mu+1}^* h_\mu
	\nonumber
\\
		&& \qquad + a^2\nu^{-1}\varepsilon^{3/2} h_{\mu+1}^* h_\mu + {\rm O}\left(\nu^{-3}\right) \left.\frac{}{}\right]
	\label{eqn:Omega0_asymp}
\end{eqnarray}

From equation~\eqnref{eqn:zetan} we can easily see that well below the driving layer,
as $s \to \infty$, the amplitude of the downward propagating sausage wave becomes
proportional to $\Omega_0 + {\cal I}_0^*$. Oddly, if we add expression~\eqnref{eqn:Omega0_asymp}
to the complex conjugate of equation~\eqnref{eqn:I0_asymp} and collect terms of
like order, all of the terms cancel to the order calculated. Therefore, for an
incident $f$ mode, in the limit of infinite frequency, the reflected wave and the
directly generated wave undergo complete destructive interference and infinite
frequency should be considered an ``excitation null".

In this limit, the driver is essentially a delta function located at the surface.
Therefore, the phase difference between the reflected wave and the directly generated
downward wave can only accrue from a phase change upon reflection off of the photosphere.
For destructive interference, this phase change must be $\pm \pi$. The actual phase
change turns out to be $\Delta\theta = -\pi$, and this can be verified asymptotically.
The phase change can be expressed using the ratio of the complex amplitudes of the upward
propagating wave and the reflected wave. These can be obtained by evaluating
equation~\eqnref{eqn:zetan} at $s=1$,

\begin{equation}
	\Delta\theta = \arg\left\{\frac{{\cal I}_0 h_\mu^*}{\Omega_0 h_\mu}\right\} \; .
\end{equation}

\noindent Since the frequency is large, the Hankel functions represented by $h_\mu$
can be expressed using their large argument expansions,

\begin{equation}
	\Delta\theta \approx \arg\left\{\frac{{\cal I}_0}{\Omega_0} e^{-4i\nu\varepsilon^{1/2}} e^{i(\mu+1/2)\pi}\right\} \; .
\end{equation}

\noindent If we now compute the ratio ${\cal I}_0/\Omega_0$ using these same large
argument expansions we find,

\begin{equation}
	\frac{{\cal I}_0}{\Omega_0} = e^{4i\nu \varepsilon^{1/2}} e^{-i(\mu+3/2)\pi} \; .
\end{equation}

\noindent Combine these last two expressions to find that $\Delta\theta = -\pi$ to leading
order.

\section{Symmetry of the Far-Field Scattering Matrix}
\label{app:SymmetryS}
\setcounter{equation}{0}

In this appendix we will demonstrate that under several commonly used boundary
conditions the far-field scattering matrix, $S_0^{n \to n'}$, is symmetric to
interchange of $n$ and $n'$ $(S_0^{n \to n'} = S_0^{n' \to n})$. We start with
equation~\eqnref{eqn:Smatrix},

\begin{equation} \label{eqn:Smatrix_D}
	S_0^{n \to n'} = \frac{i\pi}{2} \int_1^\infty ds \; s^a Q_{n'}(s) \, d_n(s)
\end{equation}

\noindent where the frequency dependence has been suppressed. First we need an
expression for the mismatch, $d_n$, which is supplied by equation~\eqnref{eqn:dn},

\begin{equation}
	d_n(s) = \frac{R(s)}{z_0^2} \left[ N_{{\rm inc}, n}(s) - N_{\parallel, n}(s) \right] \; .
\end{equation}

Using equations~\eqnref{eqn:Ninc}, \eqnref{eqn:TubeRadius} and \eqnref{eqn:Nparn}
this expression can be reduced to a sum of two terms: a term that depends solely
on the incident $p$-mode eigenfunction and a term that depends on the excited sausage
waves,

\begin{equation}
	d_n(s) = \frac{R_0^2}{2z_0^2} \, s^{-a}
		\left\{ s^\mu \left[ \frac{\mu+1}{s} \frac{dQ_n}{ds} - \frac{\lambda_n^2}{4}Q_n(s)
				+ \frac{\gamma (\beta+1)}{2\varepsilon} \frac{\nu^2}{s} Q_n(s)\right]
			-\frac{\gamma\beta}{2\varepsilon z_0} \frac{d}{ds} \left(s^\mu \zeta_{\parallel, n}\right) \right\} \; .
\end{equation}

\noindent The terms in the square brackets can be modified using the Sturm-Liouville
differential equation that the $p$-mode eigenfunctions satisfy, equation~\eqnref{eqn:sturm_liouville},

\begin{equation}
	d_n(s) = -\frac{R_0^2}{2z_0^2} \, s^{-a}
		\left[ \frac{d}{ds} \left(s^\mu \frac{dQ_n}{ds}\right) - \frac{(\gamma-2)}{2\varepsilon} \nu^2 s^{\mu-1} Q_n(s)
			+\frac{\gamma\beta}{2\varepsilon z_0} \frac{d}{ds} \left(s^\mu \zeta_{\parallel, n}\right) \right] \; .
\end{equation}
 
\noindent Insert this expression into equation~\eqnref{eqn:Smatrix_D} to obtain,

\begin{equation} \label{eqn:Smatrix_2parts}
	S_0^{n \to n'} = -\frac{i\pi}{4} \frac{R_0^2}{z_0^2} \left\{ {\rm D}_{n'n}
			+ \frac{\gamma\beta}{2\varepsilon z_0} \int_1^\infty ds \; Q_{n'}(s) \frac{d}{ds}
				\left[ s^\mu \zeta_{\parallel,n}(s)\right] \right\} \; ,
\end{equation}

\noindent where the values ${\rm D}_{n'n}$ are the elements of a differential
operator, ${\bf D}$,    

\begin{eqnarray}
	{\bf D} \, Q(s) &\equiv& \frac{d}{ds}\left[s^\mu Q(s)\right] - \frac{\gamma-2}{2\varepsilon} \nu^2  s^{\mu-1} Q(s) \; ,
\\
	{\rm D}_{n'n} &\equiv& \int_1^\infty ds \; Q_{n'}(s) {\cal\bf D} \, Q_n(s) \; .	
\end{eqnarray}

Since we have applied Hermitian boundary conditions to the $p$ modes, one can easily
verify that the operator ${\bf D}$ is also Hermitian with respect to the same set
of functions; thus, ${\rm D}_{n'n} = {\rm D}_{nn'}$. So, we now only need consider
the second term in the braces in equation~\eqnref{eqn:Smatrix_2parts}: Is this
term symmetric upon exchange of $n$ and $n'$?

The longitudinal displacement $\zeta_{\parallel,n}$ can easily be expressed using a
Green's function, $G(s,s')$,

\begin{equation} \label{eqn:GreensSolution}
	\frac{\zeta_{\parallel,n}(s)}{z_0} = -\frac{i\pi}{2} \Omega_n \psi_\parallel(s) + \int_1^\infty ds' \; G(s,s')f_n(s') \; ,
\end{equation}

\noindent where

\begin{eqnarray}
	G(s,s') &=& -\frac{i\pi}{2} s'^{(\mu+1)} \Psi(s,s') \; ,
\\
\nonumber \\
	\Psi(s,s') &\equiv&
		\left\{{\psi_\parallel(s) \, \psi_\parallel^*(s')  \qquad {\rm if} \; s'<s,
			\atop
		   	\psi_\parallel^*(s) \, \psi_\parallel(s')  \qquad {\rm if} \; s'>s,} \right.
\\
\nonumber \\
	f_n(s) &=& -\frac{\gamma(\beta+1)}{2} \frac{\nu^2}{s} \frac{dQ_n}{ds} \; .
\end{eqnarray}

\noindent This formulation is different from the one used in Appendix B because
we will exploit symmetries possessed by the Green's function and the present
formulation makes those symmetries explicit. In particular, note that the function
$\Psi(s,s')$ is symmetric upon the interchange of $s$ and $s'$.

In expression~\eqnref{eqn:GreensSolution}, $\Omega_n$ is a complex constant that
determines the boundary condition applied to the sausage wave at the surface of
the truncated polytrope. For a stress-free surface the appropriate value of $\Omega_n$
is given by,

\begin{equation} \label{eqn:BCstress-free}
	\Omega_n = i \frac{\gamma(\beta+1)}{\pi} \nu^2 \frac{{\cal Q}_n}{\cal H} - \frac{{\cal H}^*}{\cal H} {\cal I}_n \; ,
\end{equation}

\noindent and, alternatively, for an upper atmosphere with minimal reflectivity
\citep{Hindman:2008} the correct value of the boundary condition parameter $\Omega_n$ is

\begin{equation} \label{eqn:BCmax-flux}
	\Omega_n = -\frac{\gamma(\beta+1)}{2} \nu^2 h_\mu {\cal Q}_n \; .
\end{equation}

If one inserts equation~\eqnref{eqn:GreensSolution} into expression~\eqnref{eqn:Smatrix_2parts},
integrates by parts, and uses the fact that the $p$-mode eigenfunctions and the
tube wave solutions vanish at infinity ($s \to \infty$), one can obtain the following
expression after some manipulation,

\begin{equation}
	S_0^{n \to n'} = -\frac{i\pi}{4} \frac{R_0^2}{z_0^2} {\rm D}_{n'n}
		+ \frac{\pi^2 \gamma\beta}{16 \varepsilon z_0} \frac{R_0^2}{z_0^2} \left(W_{n'n} + B_{n'n}\right) \; ,
\end{equation}

\noindent where

\begin{eqnarray}
	W_{n'n} &\equiv& \frac{\gamma(\beta+1)}{2} \nu^2 \int_1^\infty \int_1^\infty ds \, ds' \; s^\mu s'^\mu \Psi(s,s')
		\frac{dQ_n'}{ds} \frac{dQ_n}{ds'} \; ,
\\
	B_{n'n} &\equiv& \frac{\Omega_n{\cal I}_{n'}}{\gamma(\beta+1)\nu^2} - {\cal Q}_{n'} \left(h_\mu\Omega_n + h_\mu^*{\cal I}_n\right) \; .
	\label{eqn:Bmatrix}
\end{eqnarray}

It is trivial to demonstrate that $W_{n'n}$ is symmetric. Since the variables $s$ and $s'$
in the integrals are dummy variables, we can exchange them ($s \to s'$ and $s' \to s$) and
then exchange the order of integration. Further, the function $\Psi(s,s')$ is symmetric to
interchange of its two arguments. Therefore, after performing these exchanges we find the
forestated symmetry, $W_{n'n} = W_{nn'}$.

Thus, the symmetry of the far-field scattering matrix hinges on the properties of the
matrix $B_{n'n}$, which contains boundary terms arising from the integration by parts
as well as a contribution that comes from the homogeneous solution appearing in the
first term in equation~\eqnref{eqn:GreensSolution}. Under some common boundary conditions
for the sausage waves, this matrix is symmetric, but in general it is not. One can readily
verify by directly inserting expressions~\eqnref{eqn:BCstress-free} or \eqnref{eqn:BCmax-flux}
into equation~\eqnref{eqn:Bmatrix}, and using the identity for the Wronskian of
two Hankel functions,

\begin{equation}
	{\cal W}\left\{H_m^{(1)}(x), H_m^{(2)}(x)\right\} = H_{m+1}^{(1)}(x)H_m^{(2)}(x) - H_m^{(1)}(x)H_{m+1}^{(2)}(x) = -\frac{4i}{\pi x} \; ,
\end{equation}

\noindent that both of the boundary conditions mentioned previously (stress free
and minimum reflectivity) produce a symmetric matrix, $B_{n'n} = B_{nn'}$. Therefore,
at least for these two boundary conditions, the far-field scattering matrix is also
symmetric.

The crucial properties that were required for this symmetry are (1) the $p$ modes
satisfy Hermitian boundary conditions, (2) the sausage wave vanishes deep in the
atmosphere ($s \to \infty$), and (3) the sausage wave satisfies a boundary condition 
at the surface of the truncated polytrope ($s = 1$) that causes the boundary term
matrix $B_{n'n}$ to be symmetric (note, not necessarily vanishing). This proof
is very reminiscent of the calculation that is performed to demonstrate that a
differential operator is Hermitian; however, in this case the operator in question
is actually an integro-differential operator (because of the integration over the
Green's function).




\def\figone{%
\begin{figure*}%
        \epsscale{1.0}%
        \plotone{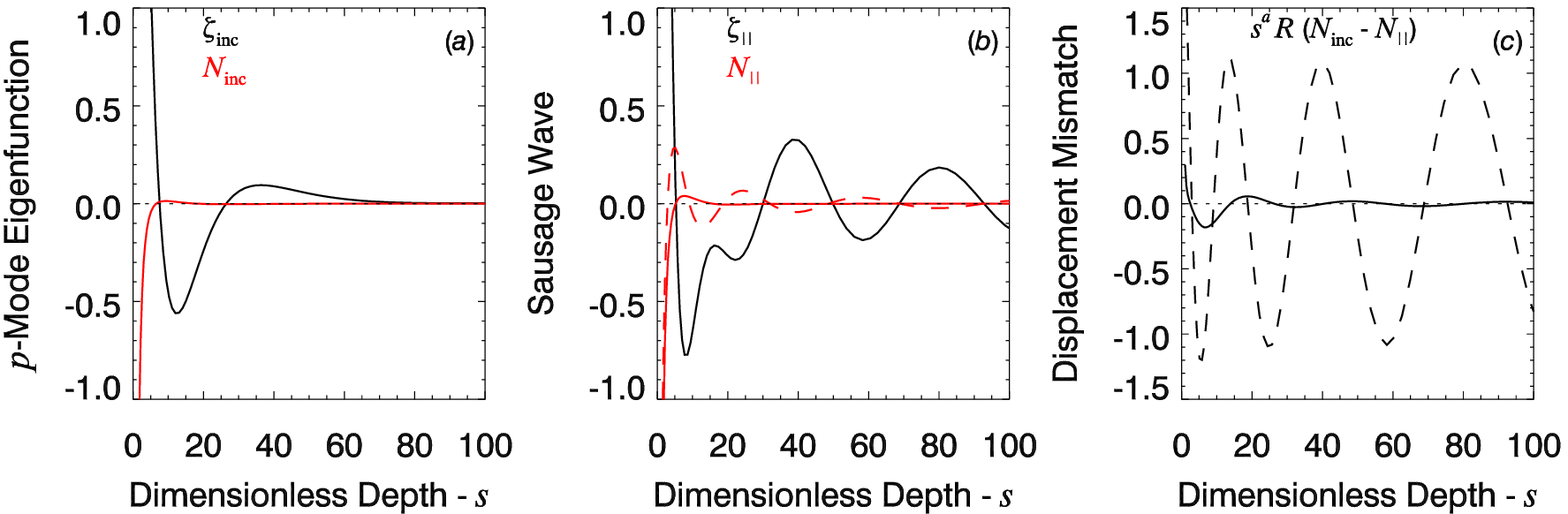}%
        \caption{\small ($a$) The vertical displacement $\zeta_{\rm inc}$ (black
curve) and the normal displacement to the flux tube's surface $N_{\rm inc}$ (red
curve) for an incident $p_2$ mode with a frequency of 3 mHz. ($b$) The vertical
and normal displacements (black and red curves, respectively) for the sausage wave
that is driven by the incident $p_2$ mode. The magnetic flux tube has a plasma
parameter of $\beta = 1$. ($c$) The resulting mismatch in the normal displacement
between the $p$ mode and the sausage wave. In all panels the solid and dashed
curves respectively represent the real and imaginary parts (when present).
\label{fig:wave_functions}}%

\end{figure*}%
}


\def\figtwo{%
\begin{figure*}%
        \epsscale{1.0}%
        \plotone{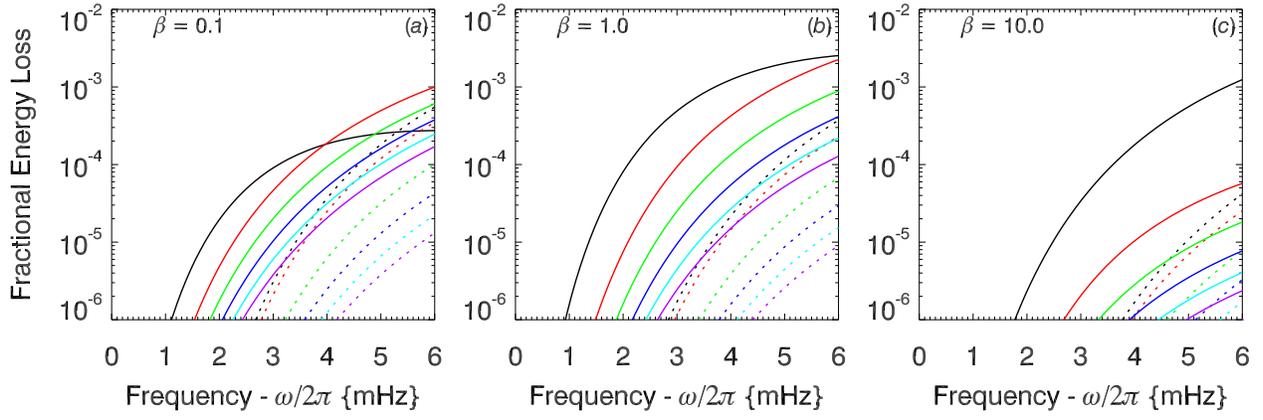}%
        \caption{\small Fractional energy lost from the incident wave as a function
of frequency for 100 km radius flux tubes with the three indicated values of the plasma
parameter $\beta$. The colors indicate the order of the incident waves ($f$: black, $p_1$:
red, $p_2$: green, $p_3$: blue, $p_4$: aqua, $p_5$: violet). The solid curves are the absorption
coefficient $\alpha_n$ computed with the scattering matrix using equation~\eqnref{eqn:Efrac_OrderR2}.
The dotted curves correspond to the energy lost from the incident mode order due to mode
mixing into all other mode orders, $M_n$, equation~\eqnref{eqn:Emodemix}.
The energy loss arising from mode mixing is generally small compared to the tube wave excitation,
although at high frequencies ($> 5$ mHz) the mode mixing can climb to one-third of the total energy
loss. Further, for the $f$ mode in the limit of large frequency, the absorption asymptotes to
zero, leading to the property that the mode mixing can dominate the energy loss. The transition
into this regime occurs when the wavelength of the sausage wave becomes much larger than the
evanescence length of the $f$ mode; thus, the transition frequency increases as $\beta$ increases.
\label{fig:absorb}}%

\end{figure*}%
}


\def\figthree{%
\begin{figure*}%
        \epsscale{1.0}%
        \plotone{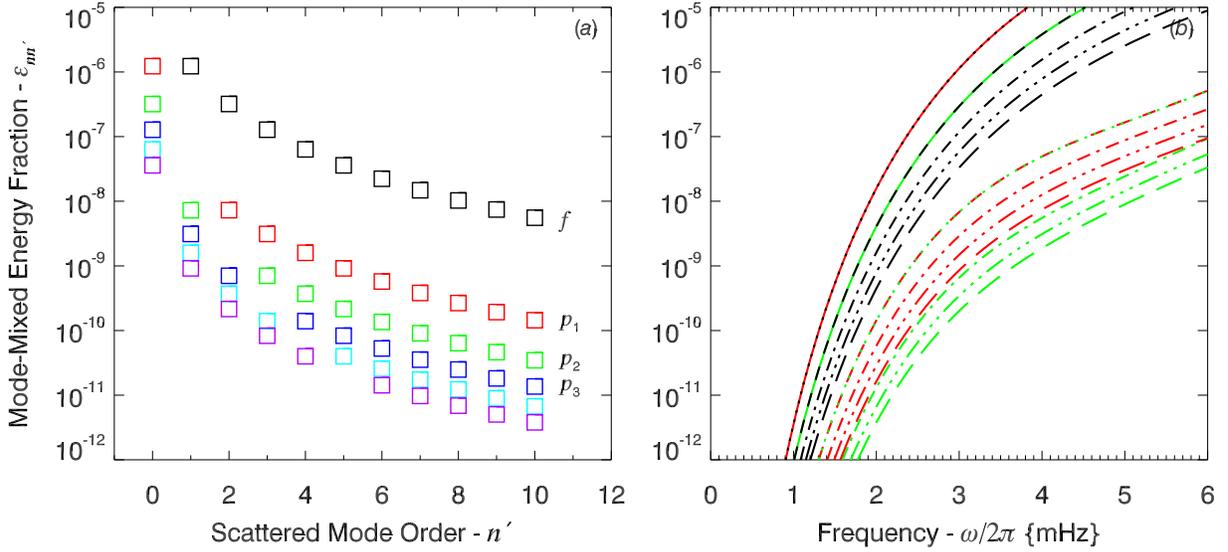}%
        \caption{\small Fractional energy lost from the incident mode due to mode mixing
induced by a magnetic tube with plasma parameter $\beta = 1$ and a radius of 100 km. ($a$)
Energy lost to each mode order of scattered waves for an incident wave of 3 mHz in frequency.
The labels and the different colors correspond to the order of the incident wave, using
the same color scheme as in Figure 2. ($b$) Energy lost to mode mixing as a function of
frequency. The colors indicate the incident mode order ($f$ through $p_2$), while the
linestyles correspond to different orders of the scattered modes ($f$: solid, $p_1$: dots,
$p_2$: short dashes, $p_3$: dot-dashed, $p_3$: three dots and a dash, $p_4$: long dashes).
In both panels careful examination verifies that the modulus of the scattering matrix
is symmetric, $|S_0^{n \to n'}| = |S_0^{n' \to n}|$.
\label{fig:modemix}}%

\end{figure*}%
}


\def\figfour{%
\begin{figure*}%
        \epsscale{1.0}%
        \plotone{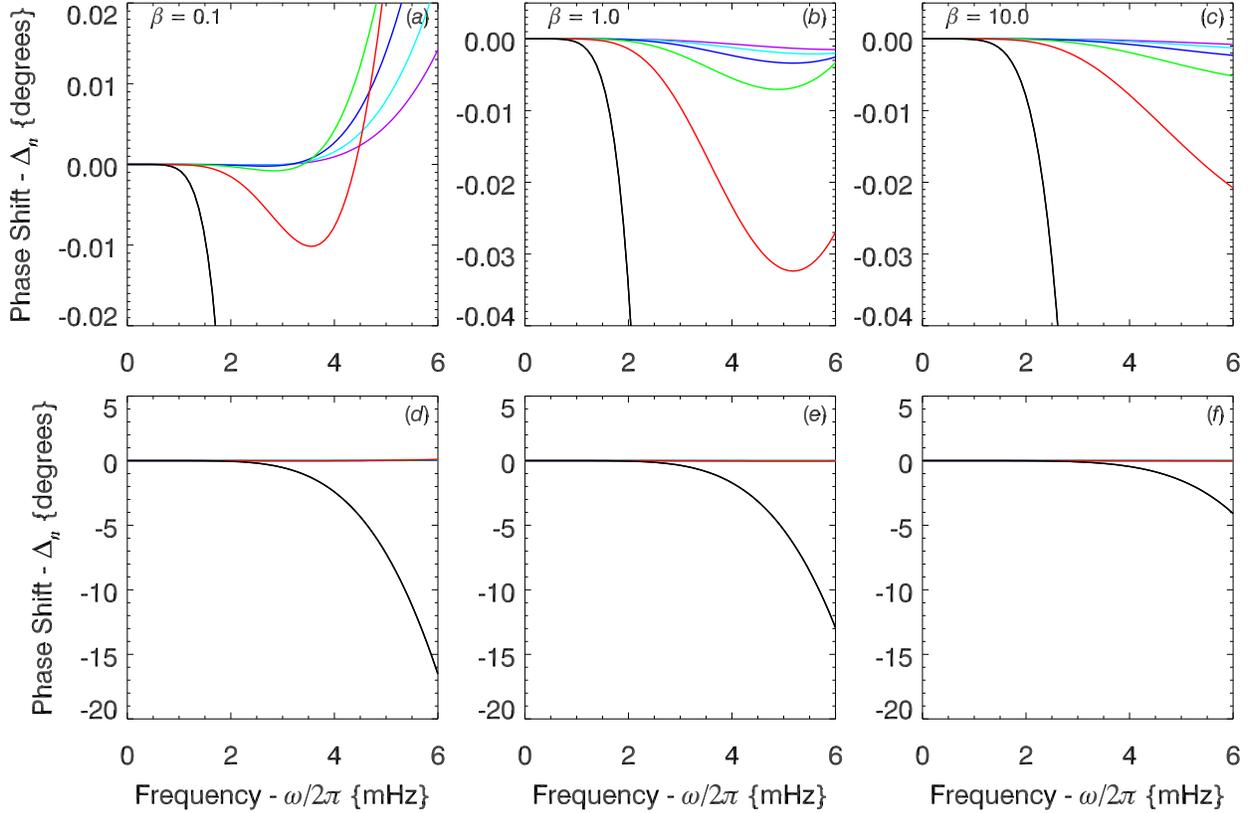}%
        \caption{\small The phase shift as a function of frequency for flux tubes
with a radius of 100 km and with the three indicated values of $\beta$. The different
colors represent different incident mode orders ($f$: black, $p_1$: red, $p_2$: green,
$p_3$: blue, $p_4$: aqua, $p_5$: violet). The bottom panels show the same quantity,
but with a differently scaled ordinate in order to better illustrate the $f$ mode.
Both positive and negative phase shifts are possible for $p$ modes and are universally
small. Whereas for the $f$ mode, the shifts are negative and substantial, reaching values
as large as -15$^\circ$.
\label{fig:phase_shift}}%

\end{figure*}%
}


\def\figfive{%
\begin{figure*}%
        \epsscale{1.0}%
        \plotone{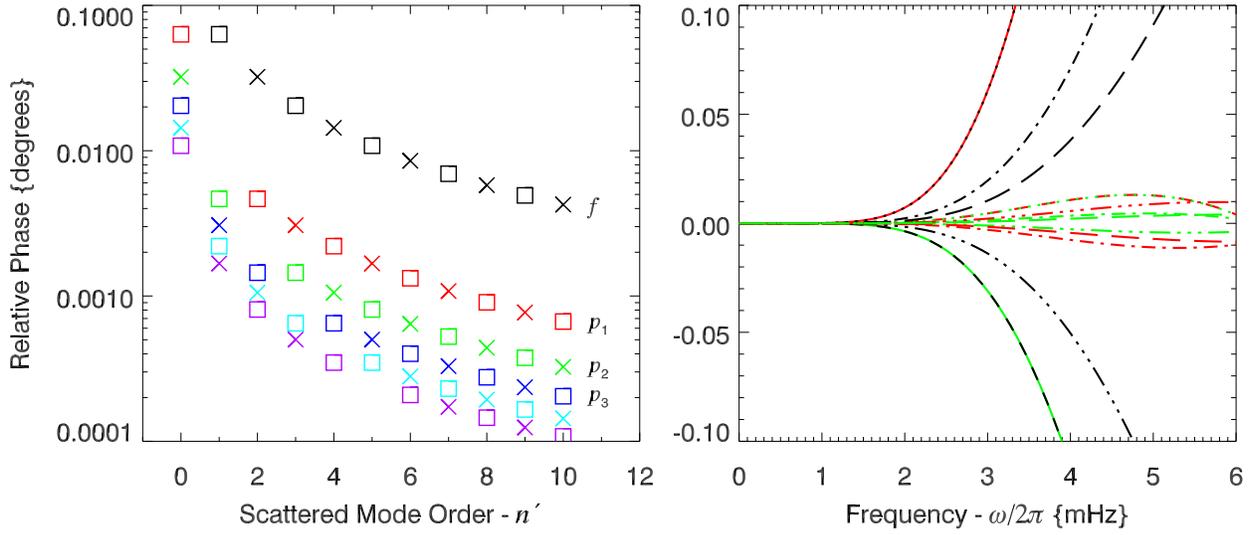}%
        \caption{\small The phase of the scattered waves measured relative to the phase
of the incident wave. ($a$) Absolute value of the phase of each mode order of scattered
waves for an incident wave of 3 mHz in frequency. The $\times$ symbols correspond to positive
phases and $\Box$ symbols to negative phases. The different colors (and labels) indicate
the order of the incident wave. ($b$) Relative phase of each scattered wave as a function
of frequency. The colors indicate the incident mode order ($f$ through $p_2$), while the
linestyles correspond to different orders of the scattered modes as in Figure 3 ($f$: solid,
$p_1$: dots, $p_2$: short dashes, $p_3$: dot-dashed, $p_3$: three dots and a dash,
$p_4$: long dashes). In both panels careful examination verifies that the phase of the
scattering matrix is symmetric, $\arg\{S_0^{n \to n'}\} = \arg\{S_0^{n' \to n}\}$.
\label{fig:scatphase}}%

\end{figure*}%
}


\def\figsix{%
\begin{figure*}%
        \epsscale{0.5}%
        \plotone{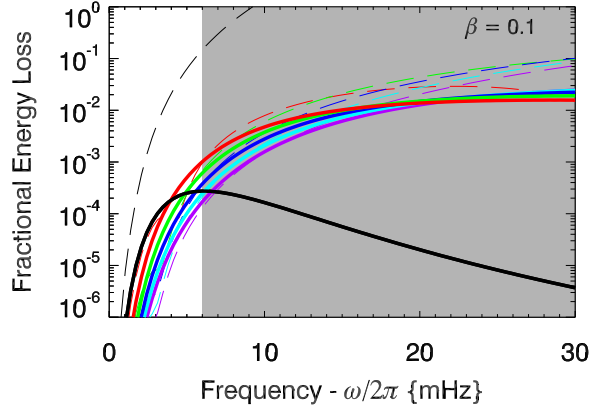}%
        \caption{\small The high-frequency behavior of the absorption profiles. The
thick solid curves are the absorption coefficient calculated via the far-field scattering
matrix with equation~\eqnref{eqn:Efrac_OrderR2}, with the different colors indicating
the order of the incident mode. The $f$ mode's high-frequency behavior is fundamentally
different from the $p$ modes because it is a surface wave.  At high frequency, driving
from the $f$ mode becomes spatially localized at the surface, wheras the driving from
$p$ modes remains spatially distributed. The thin, dashed curves show the modulus of the
scattering matrix $\left|S_0^{n \to n}(\omega)\right|$. For the perturbation scheme used
here, the scattering matrix must be a small quantity. Clearly, for the $f$ mode, this
weak-scattering approximation is invalid for frequencies exceeding roughly 6 mHz (the
shaded region). We show such large frequencies merely to illustrate the mathematical
behavior at high frequencies.
\label{fig:highfreq}}%

\end{figure*}%
}

\figone
\figtwo
\figthree
\figfour
\figfive
\figsix

\end{document}